\documentclass[twocolumn,aps,prb,showpacs]{revtex4}

\usepackage{graphicx,color}   
\usepackage{amsfonts}
\usepackage{amsmath,amssymb,mathrsfs}
\usepackage{bm}

\newcommand{\bs}[1]{\boldsymbol{#1}}

\newcommand{\sumk}{\sum_{\bs{k}}}

\newcommand{\vac}{\left|\,0\,\right\rangle}

\newcommand{\ket}[1]{\left|#1\right\rangle}
\newcommand{\bra}[1]{\left\langle#1\right|}

\newcommand{\up}{\uparrow}
\newcommand{\dw}{\downarrow}
\newcommand{\nn}{\nonumber}
\newcommand{\eps}{\varepsilon}
\newcommand{\pa}{\partial}

\def\ie{\emph{i.e.},\ }
\def\eg{\emph{e.g.}\ }
\def\ea{\emph{et~al.}}

\begin{document}
\title{Topological Insulators and Mott Physics from the Hubbard Interaction}

\author{Stephan Rachel}
\altaffiliation{Electronic address:~stephan.rachel@yale.edu}
\affiliation{Department of Physics, Yale University, New Haven, CT 06520, USA}
\author{Karyn Le Hur}
\altaffiliation{Electronic address:~karyn.lehur@yale.edu}
\affiliation{Department of Physics, Yale University, New Haven, CT 06520, USA}

\pagestyle{plain}
\begin{abstract}
We investigate the Hubbard model on the honeycomb lattice with intrinsic spin orbit interactions as a paradigm for two--dimensional topological band insulators in the presence of interactions. Applying a combination of Hartree--Fock theory, slave--rotor techniques, and topological arguments, we show that the topological band insulating phase persists up to quite strong interactions. Then we apply the slave-rotor mean-field theory and find a Mott transition at which the charge degrees of freedom become localized on the lattice sites. The spin degrees of freedom, however, are still described by the original Kane--Mele band structure. Gauge field effects in this region play an important role. When the honeycomb layer is isolated then the spin sector becomes already unstable toward an easy plane Neel order. In contrast, if the honeycomb lattice is surrounded by extra ``screening'' layers with gapless spinons, then the system will support a fractionalized topological insulator phase with gapless spinons at the edges. For large interactions, we derive an effective spin Hamiltonian.
 
\end{abstract}


\maketitle

\section{Introduction}\label{sec:intro}

Topological insulators embody a new class of topological states which have attracted great attention recently\cite{qi-10pt33,hazan-10arXiv:1002.3895,moore10n194,buttiker09s278}.
The key for this flourishing development is the understanding that spin orbit interactions can realize topological insulating phases\cite{kane-05prl146802,kane-05prl226801,fu-07prl106803,moore-07prb121306,roy09prb195322}. The theoretical prediction of such phases in real materials\cite{bernevig-06s1757,fu-07prb045302,zhang-09np438} as well as their experimental observations\cite{koenig-07s766,hsieh-08n970,hsieh-09s919,xia-09np398,chen-09s178,roushan-09n1106,hsieh-09n1101} are responsible for the success of this rapidly developing field.

A topological insulator exhibits a bulk energy gap (like ordinary insulators) while the edge (or surface in three dimensions) has gapless states which are protected by time reversal symmetry. The topological difference between a topological insulator and an ordinary band insulator is characterized by a $\mathbb{Z}_2$ invariant\cite{kane-05prl146802} which is non-zero in the topological phase. The existence of this topological quantum number as well as the quantized spin Hall conductivity inspired the field\cite{xu-06prb045322,sheng-06prl036808,qi-06prb045125,fu-06prb195312,wu-06prl106401,teo-08prb045426,lee-08prl186807,levin-09prl196803,prodan09prb125327,wang-09arXiv:0906.5118,lee-09arXiv:0908.2490,xu09arXiv:0908.2147}, in particular, it was shown that the topological insulator phase -- or in two dimensions also called {\it quantum spin Hall} (QSH) effect -- is stable against weak disorder and weak interactions\cite{xu-06prb045322,lee-08prl186807}. Inside a topological insualtor, Maxwell's laws of electromagnetism are altered by an additional topological term with a quantized coefficient, which gives rise to interesting physical effects\cite{qi-08prb195424,qi-09s1184}.

A major role has been played by a simple model introduced by Kane and Mele\cite{kane-05prl146802,kane-05prl226801} consisting of a hopping and an intrinsic spin orbit term on the honeycomb lattice. 
The Kane--Mele model (without the Rashba term) essentially consists of two copies with different sign for up and down spins of a model introduced earlier by Haldane\cite{haldane88prl2015}.
Haldane's pioneering work realizes the Quantum Hall effect without an external uniform magnetic field. It breaks, however, time reversal symmetry (necessary for the quantum Hall effect) which can be restored by taking two copies with different signs for the spins together (as Kane and Mele did). 
Originally they proposed the model as realization of the QSH effect in graphene\cite{kane-05prl146802,kane-05prl226801} and
today it should be seen as a paradigm, a perfect theoretical model for topological insulator phases in two dimensions. The honeycomb lattice is definitely interesting on its own due to the striking development within the graphene community\cite{castroneto-09rmp109} but also more exotic phenomena like zero modes\cite{ghaemi-07arXiv:0709.2626,bergman-09prb184520} have been discovered, for example. 
The honeycomb lattice has also attracted some attention
in relation with exotic phases of light and the Jaynes--Cummings
lattice model\cite{koch-09pra023811}.
The additional spin orbit interactions now make the difference and are responsible for the existence of a topological insulator phase\cite{kane-05prl146802,kane-05prl226801}. Consequently, real materials with (strong) spin orbit interactions have been attracted particular notice\cite{jackeli-09prl017205,shitade-09prl256403,pesin-09np376,yang-10arxiv:1004.4630}. It was also shown that 
strong nearest- and next-nearest neighbor repulsions can imitate the intrinsic spin orbit interactions such that QSH phases are stabilized in the absence of spin orbit coupling\cite{raghu-08prl156401,weeks-10prb085105,wen-10arXiv:1005.4061}. Beside mercury telluride quantum wells and the Kane--Mele model, topological insulating phases in two dimensions are found to exist in the Kagome lattice\cite{guo-09prb113102} and the decorated honeycomb lattice\cite{ruegg-09arxiv:0911.4722} provided the presence of spin orbit interactions.

Other aspects of topological insulators are disorder induced topological phases as predicted for the HgTe quantum wells\cite{li-09prl136806,groth-09prl196805} and for three--dimensional systems\cite{ostrovsky-09arXiv:0910.1338,biswas-09arXiv:0910.4604} and the proposed existence of axions on the surface of bismuth-tin alloys\cite{qi-08prb195424}. Axions were postulated more than 30 years ago in the context of the standard model\cite{wilczek87prl1799} and their effective action has now been recovered in topological insulators raising hope to detect this dynamical axion field experimentally. Moreover, a QSH phase in ferromagnetic graphene was predicted\cite{sun-10prl066805} which is protected by the product of charge-conjugation and time reversal symmetry. Most recently, a new family of topological insulators has been discovered\cite{lin-10arXiv:1003.0155,chadov-10arXiv:1003.0193} in ternary Heusler compounds. Their additional open $f$-shell element might be the key for the realization of exotic topological effects.

Another promising path for the realization of topological phases and, in particular, QSH phases consists of cold atomic gases loaded into optical lattices\cite{bloch-08rmp885} which are subjected by a synthetic magnetic field. Such a field has a similar effect on the neutral atoms as a magnetic field coupled to electrons and has been demonstrated experimentally\cite{lin-09n628}. A possible experiment to realize a topological insulator was proposed recently\cite{goldman-10arXiv:1002.0219,wu08prl186807,stanescu-09pra053639,stanescu-09arXiv:0912.3559}. In this spirit, a realization of a topological band insulator seems to be feasible in the near future with possibly two major advantages: (i) tuning of the topological insulator band gap or of the details of the engineered Hamiltonian and (ii) availability of onsite-interactions (Hubbard model) with tunable interaction strength.

In this paper, we investigate the Hubbard model with intrinsic spin orbit interactions on the honeycomb lattice which corresponds to the Kane--Mele (KM) model with interactions. Some aspects of the interacting KM model was studied in Refs.\,\onlinecite{goryo-09arXiv:0905.2296,goryo-10arXiv:1007.1507}. A general theory of interaction effects in topological insulators has been proposed introducing a topological order parameter in terms of the full Green's function\cite{wang-10arXiv:1004.4229}.
We consider the half--filled case at zero temperature. While the (non-interacting) KM model is known to realize a topological band insulator (TBI) phase, it is also expected that for sufficiently strong electron--electron interactions magnetic order will take place. Therefore we want to clarify what happens and which phases are present when adding interactions -- ranging from very weak to very strong. We focus on the dominant phases at finite spin orbit coupling. For very weak or no spin orbit coupling additional (spin liquid) phases might exist but are beyond the scope of this paper. First, we show that interestingly the TBI phase subsists up to quite strong interactions.

Then, applying the slave rotor mean-field procedure, we investigate the limit of stronger interactions where the charge degrees of freedom form a Mott insulator 
whereas the spin degrees of freedom are described by a renormalized KM model. In a Mott phase, adding a particle at a given site costs the Hubbard onsite energy $U$ (in contrast, excitations carry a well-defined momentum in the TBI phase). At the mean-field level, this
phase has all the properties of a spin liquid (which preserves time-reversal symmetry) with gapless spinon
excitations at the edges and is characterized by a hidden order parameter in the spin sector similar to that in the original KM model\cite{kane-05prl146802,raghu-08prl156401}. On the other hand, one should not underestimate the effect of dynamical compact U(1) gauge fields, especially in two dimensions\cite{polyakov75plb82}. 

\begin{figure}[t!]
\centering
\includegraphics[scale=1.]{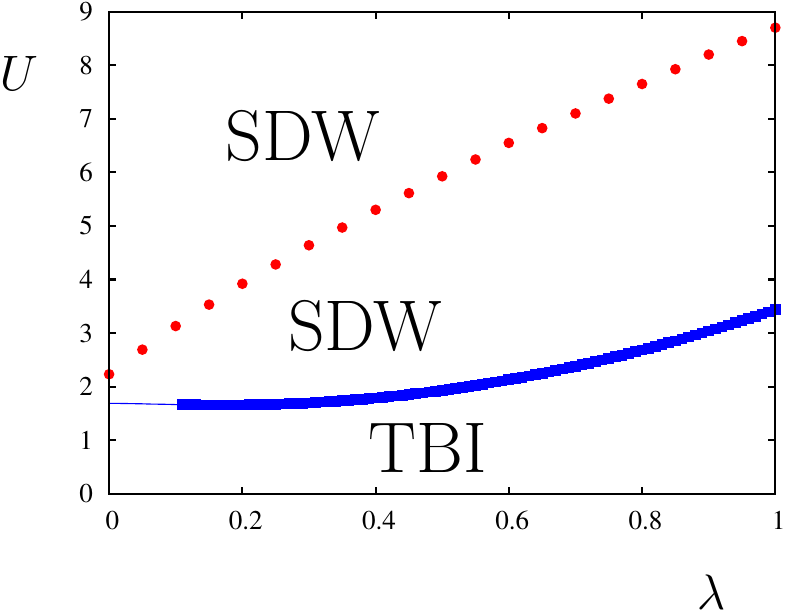}\\
\caption{(Color online) Phase diagram of the isolated honeycomb layer where the proliferation of instantons
produces a Neel order in the XY plane already in the entrance of the Mott phase. Above the red dashed line the SDW phase can be described in terms of a mean-field Hartree--Fock theory whereas below the red dashed line the easy plane Neel order emerges as a result of subtle gauge fluctuations beyond the mean-field solution. (The precise nature of the ``transition'' associated with the
red dashed line is beyond the scope of this paper.)}
\label{fig:phasedia-sr}
\end{figure}

\begin{figure}[t!]
\centering
\includegraphics[scale=1.]{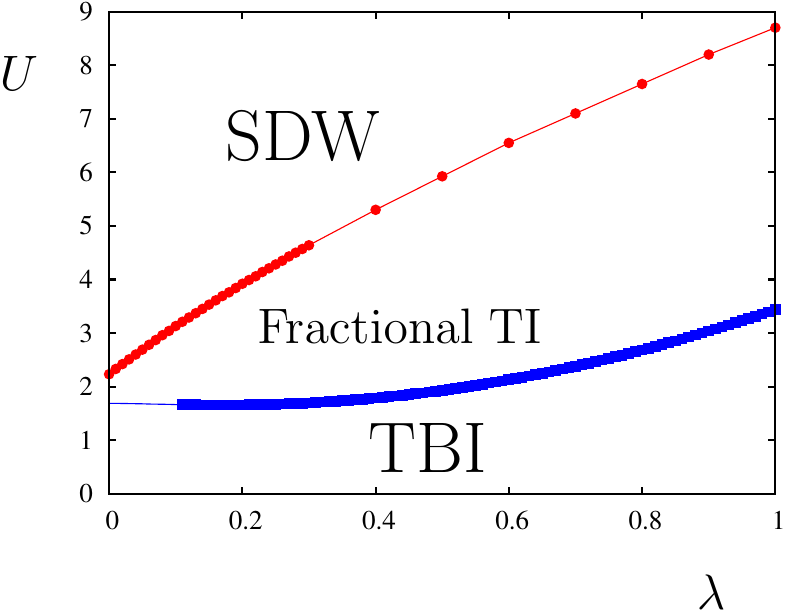}\\
\caption{(Color online) Phase diagram in the presence of additional screening layers (with gapless spinons) allowing to screen the gauge field and therefore stabilize the Fractionalized TI phase found at the mean-field level. Here, charge degrees of freedom are in the Mott regime and spin degrees of freedom form a spin liquid with gapless edge spinons.}
\label{fig:phasedia-sr1}
\end{figure}
In particular, one predicts\cite{young-08prb125316} that such a spin liquid phase (with gapless edge spinons) found at the mean-field level can only be stable beyond the mean-field limit if other gapless layers (spinons) are present to screen the gauge field and suppress the gauge fluctuations. Potential candidates can be found in Refs.\,\onlinecite{wen02prb165113,fisher08arxiv0812.2955}. Furthermore, Mott physics
will suppress the single-particle tunneling at the edges \cite{lehur01prb165110} such that the lowest relevant coupling between layers is the usual spin-spin interaction which may remain irrelevant \cite{young-08prb125316}, then preserving the gapless edge spinons. Phases exhibiting similar spin-charge separation were also reported in other systems\cite{pesin-09np376,shitade-09prl256403,young-08prb125316} and for topological insulators in the presence of a $\pi$ flux\cite{ran-08prl086801,qi-08prl086802}. In contrast, if the honeycomb layer is isolated then the proliferation of instantons will fatally result in a Neel ordering in the XY plane\cite{ran-08arxiv:0806.2321,hermele-08prb224413}. The two distinct scenarios at the Mott transition are reported in Fig.\,\ref{fig:phasedia-sr} and \ref{fig:phasedia-sr1}. Following
Ref. \onlinecite{young-08prb125316}, Fractionalized TI refers to the spin liquid type Mott phase with gapless spinons which preserves time-reversal symmetry and SDW in the two figures always refers to the occurrence of a spin density wave formed in the XY plane.

For very large interactions, applying a conventional Hartree--Fock procedure and deriving an effective spin Hamiltonian, we show that SDW phases with XY ordering are allowed on the honeycomb lattice when adding the spin-orbit term coupling next nearest neighbors. 

The paper is organized as follows. In Sec.\,\ref{sec:model} we introduce the KM model, re-derive some of its basic properties, and introduce the (Hubbard) interaction we consider throughout the paper. In Sec.\,\ref{sec:HF} we apply the Hartree Fock method in order to show that a conventional SDW phase with ordering in the XY plane appears at large $U$. In addition, we derive an effective spin model. Then, in Sec.\,\ref{sec:BHF}, we use a mean field approach in momentum space as well as the slave rotor picture to argue that the TBI phase as present in the original KM model is stable beyond renormalization group results\cite{kane-05prl226801,shankar94rmp129} up to moderate interactions. Then, we apply the slave rotor theory of Florens and Georges\cite{florens-02prb165111,florens-04prb035114,zhao-07prb195101} and discuss the intermediate region and the gauge field effects more thoroughly.

\section{Model and general considerations}\label{sec:model} 

\begin{figure}[t!]
\centering
\includegraphics[scale=1.]{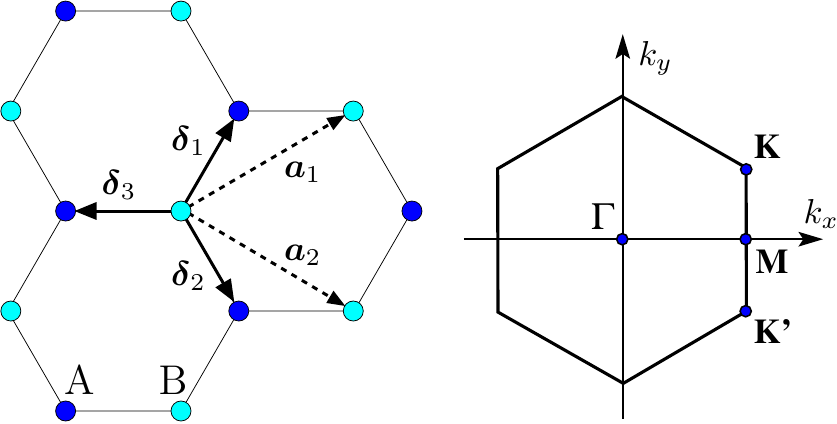}
\caption{(Color online) Left: Honeycomb lattice consisting of two interpenetrating triangular lattices, A (dark blue dots) and B (cyan dots), with its lattice vectors $\bs{a}_1$ and $\bs{a}_2$ (dashed arrows). In addition, the three nearest neighbor vectors $\bs{\delta}_i$ ($i=1,2,3$) are shown connecting the two sublattices (solid arrows). Right: Corresponding Brillouin zone with the two inequivalent Dirac cones $K$ and $K'$ and the high-symmetry points $\Gamma$ and $M$.}
\label{fig:hc-lattice}
\end{figure}

The Kane--Mele (KM) model\cite{kane-05prl146802,kane-05prl226801} which might be considered as a spinful version of the Haldane model consists of two parts, a nearest neighbor hopping term and a second neighbor hopping spin orbit term on the honeycomb lattice,
\begin{equation}\label{ham:kmh}
\mathcal{H}=-t\sum_{\langle ij \rangle}\sum_\sigma c_{i\sigma}^\dagger
c_{j\sigma}^{\phantom{\dagger}} + \, i \lambda \sum_{\ll ij \gg}
\sum_{\sigma\sigma'}
\,\nu_{ij}\,\sigma^z_{\sigma\sigma'} \,c_{i\sigma}^\dagger c_{j\sigma'}^{\phantom{\dagger}}\ .
\end{equation}
Here $c_{i\sigma}^{\phantom{\dagger}}$ is an electron annihilation operator either on sublattice A or B (then denoted by $a_{i\sigma}$ or $b_{i\sigma}$, respectively) fulfilling the fermionic standard anti-commutation relations $\{ c_{i\sigma}^{\phantom{\dagger}},c_{j\sigma'}^\dagger\}=\delta_{ij}\delta_{\sigma\sigma'}$. As usual $t$ is the hopping integral and $\lambda$ is the spin orbit coupling, $\langle ij \rangle$ denotes nearest neighbor and $\ll ij \gg$ next nearest neighbor sites, $\sigma^z$ is the third Pauli matrix and $\nu_{ij}=\pm 1$ as discussed below. 
Throughout the paper we consider the Rashba spin orbit interaction to be zero.
The lattice vectors of the honeycomb lattice are given by
\begin{equation}
\bs{a}_1 = \frac{a}{2}\big( 3,\sqrt{3}\big),\qquad
\bs{a}_2 = \frac{a}{2}\big( 3,-\sqrt{3}\big)
\end{equation}
and shown in Fig.\,\ref{fig:hc-lattice}. The lattice vectors have the length $\sqrt{3}a$ while the lattice spacing $a$ is the distance between neighboring atoms A and B. Note that our notation of the honeycomb lattice is adapted from the review of Castro\,Neto \ea\cite{castroneto-09rmp109}. We further have the nearest neighbor vectors
\begin{equation}
\bs{\delta}_1=\frac{a}{2}(1,\sqrt{3}),~~~\bs{\delta}_2=\frac{a}{2}(1,-\sqrt{3}),~~~
\bs{\delta}_3 = a(-1,0)
\end{equation}
which are also shown in Fig.\,\ref{fig:hc-lattice}. The six next-nearest neighbor vectors $\bs{\delta}_i'$ are given by $\bs{\delta}'_{1,2}=\pm \bs{a}_1$, $\bs{\delta}'_{3,4}=\pm \bs{a}_2$, and $\bs{\delta}'_{5,6}=\pm(\bs{a}_2-\bs{a}_1)$. In what follows we set $a=\hbar=1$.
Throughout the paper $N_\Lambda$ denotes the number of unit cells, while $N$ is the number of particles. Hence,  the number of lattice sites is $2N_\Lambda$ and at half filling  $N=2 N_\Lambda$. If needed, we refer to the sublattices A and B as $\Lambda_A$ and $\Lambda_B$.
The previous definitions imply
\begin{equation}
\sum_{i\in\Lambda}=\sum_{i\in\Lambda_A}=\sum_{i\in\Lambda_B}=\sum_{\bs{k}\in\,{\rm BZ}}=N_\Lambda\ .
\end{equation}

As a first step we wish to reproduce the energy bands due to nearest neighbor hopping
and switch to momentum space via 
\begin{equation}
c_{i\sigma}=\frac{1}{\sqrt{N_\Lambda}}\sum_{\bs{k}} e^{i\bs{k}\bs{R}_i}
c_{\bs{k}\sigma}
\end{equation}
which yields:
\begin{eqnarray}
\mathcal{H}_t &=& -t \sum_{\langle ij \rangle} \sum_\sigma 
\Big( a_{i\sigma}^\dagger
b_{j\sigma}^{\phantom{\dagger}} + {\rm h.c.} \Big)\\[10pt]
&=&\sum_{\bs{k}\sigma} \big( a_{\bs{k}\sigma}^\dagger, b_{\bs{k}\sigma}^\dagger\big)
\underbrace{\left(\begin{array}{cc}0&-g\\[5pt]-g^{\star}&0\end{array}\right)}_{{H}_{\bs{k}}}
\left(\begin{array}{c}a_{\bs{k}\sigma}\\[7pt] 
b_{\bs{k}\sigma}\end{array}\right) \\[10pt]
\label{diag:T0} &=&\sum_{\bs{k}\sigma} \big( l_{\bs{k}\sigma}^\dagger, u_{\bs{k}\sigma}^\dagger\big)
\left(\begin{array}{cc}-|g|&0\\[5pt] 0&|g|\end{array}\right)
\left(\begin{array}{c}l_{\bs{k}\sigma}\\[7pt] 
u_{\bs{k}\sigma}\end{array}\right)\ .
\end{eqnarray}
The function $g$ is given by $g\equiv g(\bs{k})=t\sum_{j=1}^3 e^{i\bs{k}\bs{\delta}_j}$.
Here we used the unitary transformation matrix
\begin{equation}\label{trafo:gamma=0}
T_0=\left(\begin{array}{cc} \frac{1}{\sqrt{2}}\frac{g}{|g|}&\frac{-1}{\sqrt{2}}\frac{g}{|g|} \\[10pt]\frac{1}{\sqrt{2}} & \frac{1}{\sqrt{2}} \end{array} \right)
\end{equation}
to diagonalize ${H}_{\bs{k}}$ via $T_0^\dagger {H}_{\bs{k}} T_0={\rm diag}(-|g|,|g|)$.
Calculating $\pm |g|$ explicitly results in the well known tight binding spectrum of the honeycomb lattice
\begin{equation}\begin{split}
&E(\bs{k})=\pm | g |=\\[5pt]
&~~\pm t\left[3+2 \cos{(\sqrt{3}k_y)}+
4\cos{(\sqrt{3}k_y/2)}\cos{(3 k_x/2)}\right]^{\frac{1}{2}}\ ,
\end{split}\end{equation}
two particle hole symmetric bands which touch each other at the six corners of the Brillouin zone (BZ) corresponding to two inequivalent points. Expanding around these special points reveals a linear dispersion which gives rise to the name Dirac points. The positions of two inequivalent Dirac points which are shown in Fig.\,\ref{fig:hc-lattice} are 
\begin{equation}
\bs{K}=\left(\frac{2\pi}{3},\frac{2\pi}{3\sqrt{3}}\right),~~~\bs{K}'=\left(\frac{2\pi}{3},-\frac{2\pi}{3\sqrt{3}}\right)\ .
\end{equation}
Although it is very convenient to expand around the Dirac points and formulate a Dirac equation on the honeycomb lattice we will keep throughout the paper
the full tight-binding model. 

As a second step, we consider the intrinsic spin orbit term\cite{kane-05prl226801} of the KM Hamiltonian \eqref{ham:kmh}. The expression $\nu_{ij}$ gives $\pm 1$ depending on the orientation of the sites. A formal definition is
\begin{equation}
\nu_{ij} = \left( \hat{\bs{d}}_1 \times \hat{\bs{d}}_2 \right)_z
\end{equation}
where $\hat{\bs{d}}_1$ and $\hat{\bs{d}}_2$ are the unit vectors connecting the sites $j$ and $i$. Essentially, making a left turn yields ``$-1$'' while a right turn ``$+1$''. 
Note that hopping  from a site of sublattice A in direction $\bs{\delta}'_j$ would yield 
the opposite sign than hopping from a site of sublattice B in the same direction.
Hence we should keep in mind that $\nu_{ij} \propto \tau^z$ (where $\tau^z$ is again the third Pauli matrix).
\begin{figure}[t!]
\begin{center}
\includegraphics[scale=1.]{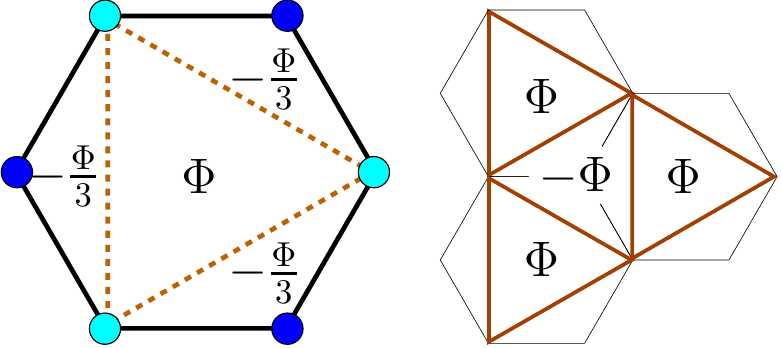}
\caption{(Color online) Left: Flux configuration per spin and sublattice associated with the intrinsic spin orbit term. Right: The flux configuration of one sublattice corresponds to a staggered triangular flux lattice.}
\label{fig:flux-config}
\end{center}
\end{figure}
As we will see below the spin orbit term opens a gap in the bulk. For completeness, we notice that other possible terms which open a gap in the spectrum are different from the spin orbit term. Such other terms, like a staggered sublattice potential  $H_{\rm st}=\sum_i\xi_i c_{i\sigma}^\dagger c_{i\sigma}^{\phantom{\dagger}}$ where $\xi_i=1$ on sublattice A and $\xi_i=-1$ on sublattice B, result in an ordinary band insulator and not in a topological phase since the gap is spin-independent. The spin orbit term preserves the original unit cell. We have shown the corresponding flux configuration per spin and per sublattice in Fig.\,\ref{fig:flux-config}. The net magnetic flux through a plaquette is zero following Haldane's idea\cite{haldane88prl2015}.
The flux pattern for one of the sublattices corresponds to a triangular staggered flux lattice.
Transforming the spin orbit term to momentum space leads to:
\begin{eqnarray}
\nn \mathcal{H}_{\rm SO} &=& i\,\lambda \sum_{\ll ij \gg} \sum_{\sigma\sigma'}
\nu_{ij} \sigma^z_{\sigma\sigma'} \Big( a^{\dagger}_{i\sigma}a^{\phantom{\dagger}}_{j\sigma'} + b^{\dagger}_{i\sigma}b^{\phantom{\dagger}}_{j\sigma'} \Big)\\[10pt]
\nn &=&2\lambda  \sum_{\sigma\sigma'}\sum_{\bs{k}}
\sigma^z_{\sigma\sigma'}\Big( a^{\dagger}_{\bs{k}\sigma}a^{\phantom{\dagger}}_{\bs{k}\sigma'} - b^{\dagger}_{\bs{k}\sigma}b^{\phantom{\dagger}}_{\bs{k}\sigma'} \Big)\\[5pt]
\nn &&
~\times\underbrace{\Big( - \sin{(\sqrt{3}k_y)} + 2 \cos{(3 k_x/2)}\sin{(\sqrt{3}k_y/2)}\Big)
}_{\equiv \gamma/2\lambda}\\[10pt]
&=&\sum_{\bs{k}} \Psi_{\bs{k}}^\dagger \gamma(\bs{k})\sigma^z\tau^z \Psi_{\bs{k}}\ ,
\end{eqnarray}
where $\Psi_{\bs{k}}^\dagger=(a_{\bs{k}\up}^\dagger, b_{\bs{k}\up}^\dagger,
a_{\bs{k}\dw}^\dagger, b_{\bs{k}\dw}^\dagger)$ and $\sigma^z\tau^z$ is understood as a $4\times 4$ matrix, $\sigma^z\tau^z={\rm diag}(1,-1,-1,1)$ ($\sigma^z$ for spin and $\tau^z$ for sublattices).

\begin{figure}[t!]
\centering
\includegraphics[scale=0.95]{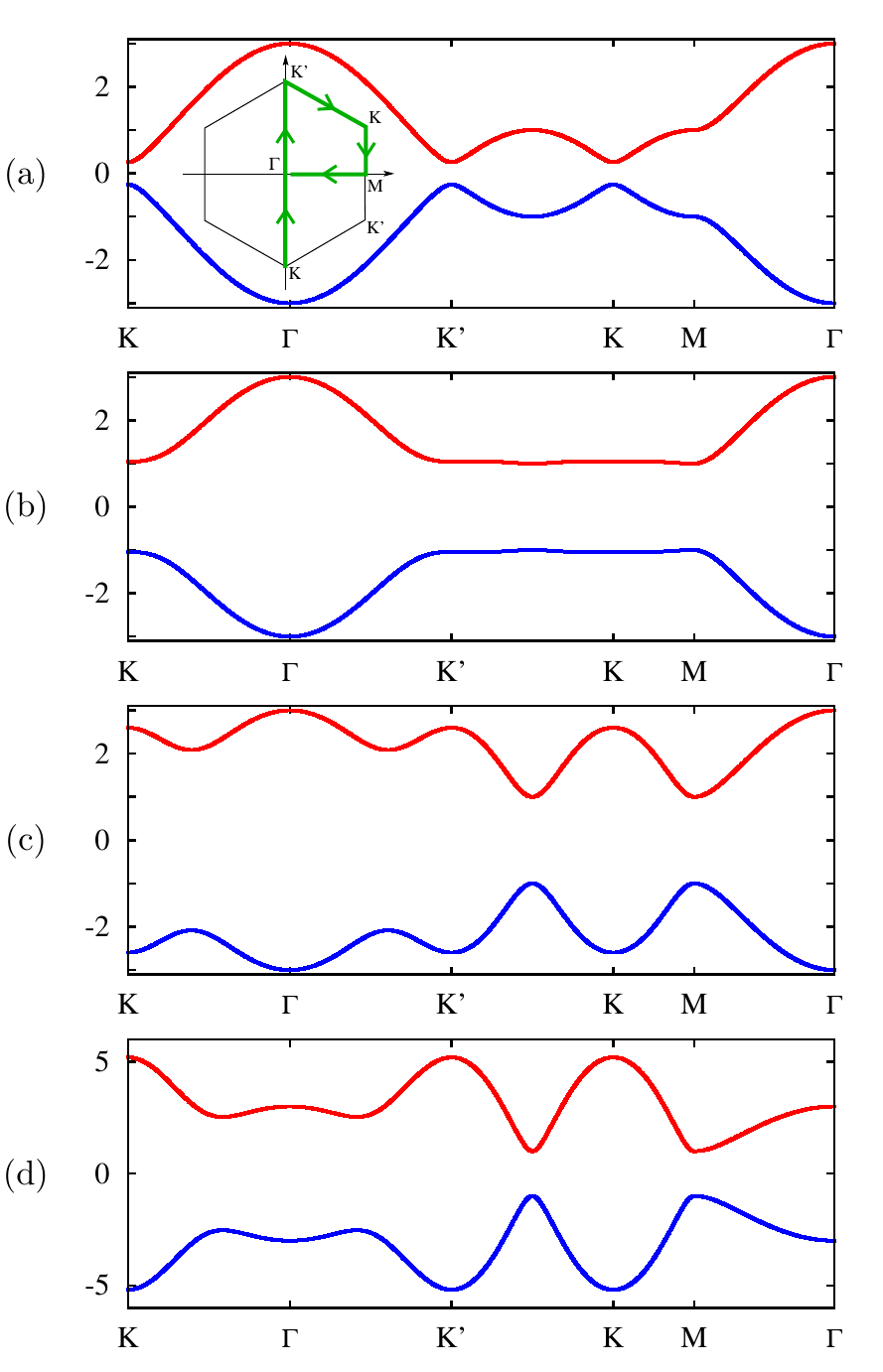}
\caption{(Color online) Energy bands ($t=1$) of the Kane--Mele model for (a) $\lambda=0.05$, (b) $\lambda=0.2$, (c) $\lambda=0.5$, and (d) $\lambda=1.0$. The ``path'' through the Brillouin zone is taken as shown in the inset of (a).}
\label{fig:energybands}
\end{figure}

The bands of the KM model are now obtained by diagonalizing the $4\times 4$ matrix of $\mathcal{H}=\mathcal{H}_t + \mathcal{H}_{\rm SO}=
\sum_{\bs{k}} \Psi_{\bs{k}}^\dagger \mathcal{H}_{\bs{k}} \Psi_{\bs{k}}$ with the matrix
\begin{equation}
\mathcal{H}_{\bs{k}}=
\left(\begin{array}{cccc}
\gamma &-g&\\[5pt]
-g^\star&-\gamma& \\[5pt] &&-\gamma &-g \\[5pt] &&-g^\star&\gamma
\end{array}\right)\ .
\label{ham:kmh-kspace}
\end{equation}
Blank entries should be thought as zeros.
As the matrix consists of two independent $2\times 2$ matrices, the diagonalization procedure is identical to Eq.\,\eqref{diag:T0} when replacing $T_0$ by $T_\up$ and $T_\dw$. The exact form of the transformation matrices, $T_\up$ and $T_\dw$, is explicitly given in Sec.\,\ref{sec:BHF}.
The single particle spectrum of the KM model in the infinite system consists of two two-fold degenerate energy bands (reflecting the Kramers degeneracy),
\begin{equation}\label{kmh-spec}
E_\pm=\pm\varepsilon(\bs{k})=\pm\sqrt{|g|^2+\gamma^2}\ ,
\end{equation}
which are plotted for several values of $\lambda$ in Fig.\,\ref{fig:energybands}. 
The spectrum in Eq.\,\eqref{kmh-spec} is still particle-hole symmetric.
An important feature is that an infinitesimal value of $\lambda$ opens an infinitesimal gap at the Dirac points.
For evaluating the gap size due to the spin orbit term we know that only the Dirac points $\bs{K}$, $\bs{K}'$ as well as the zero-energy lines in $\gamma$ play a role and it is, hence, sufficient to consider these special points. 
At the Dirac points, we find
\begin{equation}
\eps(\bs{K})=\eps(\bs{K}')=3\sqrt{3}|\lambda|\ .
\end{equation}
At the zero-energy lines of $\gamma$, we find (without loss of generality we consider here the line $k_y=0$ only)
\begin{equation}
\eps(k_x,0)=\sqrt{t^2 \left( 5+4\cos{\left( 3 k_x/2 \right)}\right)} \geq |t|\ .
\end{equation}
The minimal value $t$ is reached at the $M$ point of the Brillouin zone.
We summarize that the dispersion grows linearly with $\lambda$ at the Dirac points, but immediately when the value of $\lambda=1/(3\sqrt{3})\,t$ is reached, the gap remains constant with a gap size
\begin{equation}\label{kmh:gap-size}
\Delta=2t\quad \left( \lambda \ge 0.193\,t \right)\ .
\end{equation}

We further consider an ordinary next-nearest neighbor hopping term without spin-orbit interaction for reasons which will become clear in Sec.\,\ref{sec:SR}. 
This term is identical to $\mathcal{H}_{\rm SO}$ when omitting $i$, $\lambda$, and $\nu_{ij}$ and replacing $\sigma^z_{\sigma\sigma'}$ by $\delta_{\sigma\sigma'}$. Hence we will find a function $g_2$ instead of $\gamma$,
\begin{equation}\label{g2}
g_2(\bs{k}) = 2 \cos{(\sqrt{3}k_y)}+4\cos{(\sqrt{3} k_y/2)}\cos{(3 k_x/2)}\ .
\end{equation}
In contrast to the spin orbit term, it breaks particle-hole symmetry but does not open a gap at the Dirac points. As already mentioned this term is not present in the KM model \eqref{ham:kmh} but will become relevant in Sec.\,\ref{sec:SR}.

The aim of this paper is to investigate the effect of a local Hubbard interaction to the KM model: the Hubbard term reads
\begin{equation}\label{hubbard1}
\mathcal{H}_I' = U \sum_i n_{i\up} n_{i\dw}\ .
\end{equation}
We consider the case of half filling which allows us in principal to rewrite the Hubbard interaction as
\begin{equation}\label{hubbard2}
\mathcal{H}_I = \frac{U}{2} \sum_i \left( \sum_\sigma n_{i\sigma} - 1 \right)^2\ .
\end{equation}
This is a particularly convenient formulation for the slave rotor theory, we will, however, use the form \eqref{hubbard1} as well. Note that Eqs.\,\eqref{hubbard1} and \eqref{hubbard2} are identical at half filling.

\section{SDW phase at large $U$}\label{sec:HF}
\subsection{Mean field arguments}

In the past, a transition from the semi-metal (Dirac liquid) to a SDW phase has been evidenced in the context of the ordinary Hubbard model ($\lambda=0$) on the honeycomb lattice, when increasing the strength of the onsite interaction.
Within the Hartree--Fock approach, this transition takes place\cite{sorella-96epl699} at
$\tilde U_c=2.23\,t$.
Within quantum Monte Carlo (QMC) the transition was found\cite{sorella-96epl699} at $\tilde U_c\approx 4.5 - 5\,t$, and by means of dynamical mean field theory (DMFT) the transition\cite{jafari09epjb537} occurs even for $\tilde U_c > 10\,t$. While the critical value of the interaction parameter strongly depends on the used method there is no doubt about the existence of the SDW phase for strong interactions. The reason for the occurrence of a SDW phase is simply the bipartite nature of the honeycomb lattice.

In this Section, we will apply the Hartree Fock method to determine the phase boundary $\tilde U_c(\lambda)$ at which it becomes favorable to decouple the Hubbard interaction \eqref{hubbard1} in terms of the  sublattice magnetizations $m_i^{A}=\langle  a_{i\up}^\dagger a_{i\up}^{\phantom{\dagger}} - a_{i\dw}^\dagger a_{i\dw}^{\phantom{\dagger}} \rangle$ and $m_i^{B}=\langle  b_{i\up}^\dagger b_{i\up}^{\phantom{\dagger}} - b_{i\dw}^\dagger b_{i\dw}^{\phantom{\dagger}} \rangle$.
\begin{equation}
\begin{split}
\mathcal{H}_I' =& 
\frac{U}{4}\sum_{i\in\Lambda_A\cup\Lambda_B}
\left( (n_{i\up}+n_{i\dw})^2 - (n_{i\up}-n_{i\dw})^2 \right) \\[5pt]
\approx &\,U \sum_{i\in\Lambda_A\cup\Lambda_B} \left( \frac{1}{4} n_i^2 - \frac{1}{2} 
m_i(n_{i\up}  -n_{i\dw}) + \frac{1}{4}m_i^2  \right) \\[5pt]
=& \frac{U}{2}\sum_{i\in\Lambda} 
\Big( -m^A (n^a_{i\up} - 
n^a_{i\dw}) - m^B (n^b_{i\up} - n^b_{i\dw}) \Big)\\
& \quad\quad\quad +\frac{UN_\Lambda}{4}\Big({m^A}^2+{m^B}^2\Big) +c\ ,
\end{split}
\end{equation}
where $n_i = n_{i\up}+n_{i\dw}$, $n_{i\sigma}^{a}=a_{i\sigma}^\dagger a_{i\sigma}^{\phantom{\dagger}}$, and $U\sum_{i} n_i^2/4 = c$ is a constant in the SDW phase.
The SDW phase at large $U$ implies a Mott insulating phase for which we can evaluate $\langle n_i^2 \rangle$ explicitly. Indeed, for $U\rightarrow +\infty$, the Mott state is described by the exact wavefunction
\begin{equation}
\ket{M} = \prod_{i\in\Lambda} a_{i\sigma}^\dagger b_{i\bar\sigma}^\dagger \vac\ ,
\end{equation}
where $\sigma$ is either $\up$ or $\dw$ while $\bar\sigma$ ``points'' in the opposite direction.
We find
\begin{equation}
\bra{M}n_{i\up}^2+n_{i\dw}^2+2n_{i\up}n_{i\dw}\ket{M} = 1\ .
\end{equation}
Only one of the first two terms contributes depending on the sublattice site $i$ belongs to; the third term is always zero due to the definition of $\ket{M}$. Hence $\langle n_i^2 \rangle =1$ and $c$ is a constant in the SDW phase.
We further assume that $m_i^{(A/B)} = m^{(A/B)}$. 
In momentum space the mean field (MF) decoupled Hubbard interaction reads
\begin{equation}
\begin{split}
\mathcal{H}_I' =&\,\sum_{\bs{k}} \frac{U}{2} 
\Big( - m^A n^a_{\bs{k}\up} + m^A n^a_{\bs{k}\dw}  - m^B n^b_{\bs{k}\up} + 
m^B n^b_{\bs{k}\dw} \Big)\\
&\qquad\quad  + \frac{UN_\Lambda}{4}\Big({m^A}^2+{m^B}^2\Big) +c\ .
\end{split}\label{hubbard:decoupled-mom}
\end{equation}
While we could keep $m^A$ and $m^B$ independently, here we will search only for an antiferromagnetic solution. To be more precise we are considering only an Ising-like order parameter. In principal, one could also treat the full spin-rotational problem (\eg within a $\sigma$ model representation 
), we expect, however, no fundamental discrepancies with the simpler approach used here. In order to find the SDW phase we set
\begin{equation}
m\equiv m^A=-m^B\ .
\end{equation}
Hence the mean field Hamiltonian can be written as
\begin{equation}
\begin{split}
\mathcal{H}^{\rm HF} =& \sum_{\bs{k}} \Psi_{\bs{k}}^\dagger
\big( \mathcal{H}_{\bs{k}} + Um/2~{\rm diag}(-1,1,1,-1) \big) \Psi_{\bs{k}} \\[2pt]
&\qquad\quad + \frac{U}{2}N_\Lambda m^2\ .
\end{split}
\end{equation}
We notice that the mean field Hamiltonian coincides with the original KM model when replacing $\gamma$ by $\gamma - Um/2$ up to additional constants. Now we write the mean field free energy at $T=0$ as
\begin{equation}\label{F}
F(m) =  \sum_{\bs{k}} \Big( -2\sqrt{|g|^2 + \left(\gamma - Um/2 \right)^2}  \Big) + \frac{Um^2 N_\Lambda}{2}\ .
\end{equation}
Minimizing the free-energy with respect to $m$ yields the following self-consistent equation,
\begin{equation}\label{hf-eq}
m = \frac{1}{N_\Lambda}\sum_{\bs{k}} \frac{mU/2 - \gamma}{\sqrt{|g|^2 +\left( \gamma - m U/2\right)^2}}\ .
\end{equation}
\begin{figure}[t!]
\centering
\includegraphics[scale=1.]{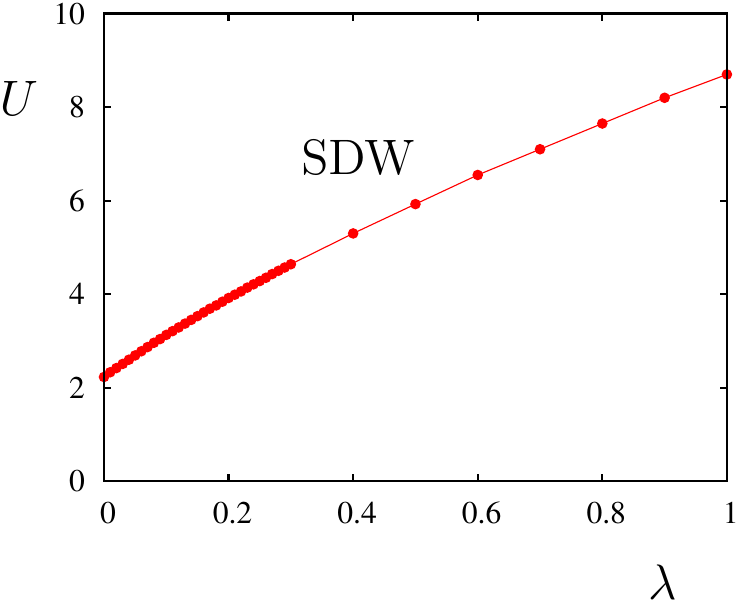}
\caption{(Color online) The numerical solution of Eq.\,\eqref{hf-eq} is shown ($t=1$). For $\lambda=0$, we confirm that the SDW transition occurs at $\tilde U_c=2.23\,t$ in agreement with the result of Sorella and Tosatti\,\cite{sorella-96epl699}. With increasing $\lambda$, $\tilde U_c$ increases up to 8.55 at $\lambda=1.0$.
The fact that $\tilde U_c(\lambda)$
increases at finite $\lambda$ may be understood from the effective spin model.}
\label{fig:hf-line}
\end{figure}
The solution of Eq.\,\eqref{hf-eq} provides the phase boundary as  shown in Fig.\,\ref{fig:hf-line}. For $\lambda=0$ we reproduce the earlier result by Sorella and 
Tosatti\cite{sorella-96epl699} which is $\tilde U_c=2.23\,t$. With increasing $\lambda$ we find that $\tilde U_c(\lambda)$ increases. The reader may notice that formally we should add a Fermi function in Eqs.\,\eqref{F} and \eqref{hf-eq} since it may play a role for larger values of $m$. Here we focus on the phase boundary only which implies small values of $m$ and neglect this point.

For the usual Hubbard model (\ie $\lambda=0$) on the honeycomb lattice the occurrence of the SDW phase is not surprising since we know that on any bipartite lattice an antiferromagnetic insulator, the SDW phase, will be favored in the limit of large $U$ (at least at half filling). The effective Hamiltonian which describes the low-energy behavior of the Hubbard model for large values of $U/t$ is the antiferromagnetic Heisenberg model. It is, however, rather unclear whether the Mott transition, \ie the phase transition from a semi metal into a gapped insulator phase, occurs simultaneously with the transition from a disordered spin state into the SDW phase. This unclearness is reflected in a current debate\cite{lee-05prl036403,herbut06prl146401,hermele07prb035125,feldner-09arXiv:0910.5360,meng-10n847,sun-09arxiv0911.3002}. 

While
in the early works\cite{sorella-96epl699} no indication was found for two separate phase transitions (Mott-Hubbard and magnetic phase transition), Lee and Lee reported a possible realization of the nodal spin liquid\cite{lee-05prl036403} directly at the Mott transition. While Herbut  favored a direct semi metal--SDW transition using a large $N$ approach\cite{herbut06prl146401}, Hermele\cite{hermele07prb035125}
 proposed the stability of the SU(2) algebraic spin liquid in a small region followed by a valence bond solid phase. A recent QMC investigation\cite{meng-10n847} rather predicts the existence of a resonating valence bond (RVB) phase between the Dirac-liquid and the SDW phase.
 
Within the Hartree Fock procedure presented above we cannot answer the question, since the SDW order already implies the Mott phase and does not tell anything about the question where the phase transition into the Mott phase or in a spin liquid phase occurs. It is, however, clear that a possible spin liquid phase must be somewhere below the transition $\tilde U_c=2.23\,t$. The SDW phase is definitely the upper boundary of such a scenario.
The fact that $\tilde U_c(\lambda)$
increases at finite $\lambda$ may be understood from the effective spin model which we will discuss now. 

\subsection{Effective spin model}
In what follows we will investigate the behavior of the spin orbit term in the strong coupling limit $U\to\infty$ with $t$ and $\lambda$ keeping fixed. Similarly to the usual Hubbard model, we expand the Hamiltonian in powers of $t/U$. The spin model can be derived in a systematic way as shown \eg in Ref.\,\onlinecite{auerbach94}, but essentially we have to consider the second order process of the Hamiltonian $\mathcal{H}_{\rm SO}=\sum_{\ll ij \gg}\left(\mathcal{H}_{\rm SO}\right)_{ij}$ with the additional prefactor $-2/U$. The minus sign respects second order perturbation theory which always lowers the energy:
\begin{equation}\label{calc:spinmodel}
\begin{split}
&\qquad \left( \mathcal{H}_{\rm SO} \right)_{ij}\left(\mathcal{H}_{\rm SO}\right)_{ji}  \\[10pt]
&= -  \nu_{ij}\nu_{ji} \lambda^2 
\left( a_{i\up}^\dagger a_{j\up}^{\phantom{\dagger}} - a_{i\dw}^\dagger a_{j\dw}^{\phantom{\dagger}} \right)
\left( a_{j\up}^\dagger a_{i\up}^{\phantom{\dagger}} - a_{j\dw}^\dagger a_{i\dw}^{\phantom{\dagger}} \right)\\[10pt]
%
&=  \lambda^2 \left( 2S_i^xS_j^x + 2 S_i^yS_j^y - 2 S_i^zS_j^z 
-\frac{1}{2}n_i n_j +n_i \right)\ .
\end{split}
\end{equation}
Without loss of generality we have considered the hopping process on sublattice A, but there is no difference with the equivalent process on sublattice B.
At half filling and for $U\to\infty$, we can assume $n_i\equiv n_{i\up}+n_{i\dw}=1$ and neglect the last terms which are constant.
Together with the mentioned factor $-2/U$ we find the effective spin model
\begin{equation}
\tilde{\mathcal{H}} = |J_2| \left( - S_i^xS_j^x - S_i^yS_j^y +  S_i^zS_j^z \right)
\end{equation}
with the exchange coupling $J_2=4\lambda^2/U$ on a triangular lattice (where we assume that the sum counts every nearest neighbor pair only once).
The spin model then consists of a XY term which favors ferromagnetic order and a Z term which favors antiparallel alignment of the spins. From the nearest neighbor hopping term we obtain an isotropic antiferromagnetic Heisenberg term with $J_1=4t^2/U$. This term stabilizes the antiferromagnetic order, \ie the SDW. The  $J_2$ term (more precisely, its $z$-component) competes with the $J_1$ term. This tends to explain the increase of the critical interaction to reach the SDW order in the phase diagram Fig.\,\ref{fig:hf-line}. 

On the other hand, the $xy$-component of the $J_2$ term favors ferromagnetic order in the $XY$ plane on each sublattice. While the ordering vector might point in any direction when $\lambda=0$, we assume that ordering within the XY plane is preferred immediately when $\lambda\not=0$. This can also be seen from energetic arguments. Once the ground state is XY ordered, ``$\up$'' and ``$\dw$'' in $\ket{M}$ refer to the $x$-component of spin. In other words, $S^z_i$ acts like a spin flip operator. Consequently, any operator containing $S^z_i$ has a zero expectation value; it particularly implies that for $U\rightarrow +\infty$, $\bra{M}\sum_k\gamma\Psi_{\bs{k}}^\dagger \sigma^z \tau^z \Psi_{\bs{k}}\ket{M}=0$.

This suggests that the SDW phase (with preferable XY order) might even persist for $J_2>J_1$ which is beyond the scope of this paper. We conclude this part by making a brief comparison with the ordinary $J_1$--$J_2$ model on the honeycomb lattice with $J_{1/2}>0$. For weak values of $J_2/J_1$, a N{\'e}el order is present. For moderate values of $J_2/J_1\approx 0.3$ a resonating valence bond (RVB) phase was proposed\cite{fouet-01ejpb241}. For stronger frustrations, a valence bond crystal exhibiting a spin gap is reasonable\cite{fouet-01ejpb241}. 
%

\section{Stability of TBI phase}\label{sec:BHF}

In this Section, we provide several arguments establishing that the topological band insulator phase present for $\lambda>0$ in the original KM model is stable (against Mott physics) not only for weak interactions\cite{kane-05prl226801,xu-06prb045322,moore-07prb121306,lee-08prl186807} but also for moderate interactions $U\sim t$.
We first consider a mean field approach in momentum space which shows that the effect of $U<2\,t$ for $\lambda> 0.2\,t$ does not affect the insulator phase and, hence, should be irrelevant for the topological band insulator phase. Then we introduce the slave rotor theory of Florens and Georges\cite{florens-02prb165111,florens-04prb035114} and argue that this provides a rigorous proof concerning the stability of the TBI phase beyond the perturbative regime. We will discuss both approaches to show that the obtained results are valid beyond the renormalization group method\cite{kane-05prl226801}. In addition, we will briefly discuss the $\mathbb{Z}_2$ topological invariant for the present situation.

\subsection{Mean field arguments}
The tight binding approach starts -- after Fourier transformation -- with the momentum operators for both sublattices ($a_{\bs{k}\sigma}$, $b_{\bs{k}\sigma}$). Diagonalization of the tight binding matrix $\mathcal{H}_{\bs{k}}$ introduces a new set of operators associated with the bands. As the spin remains a good quantum number (see \eg Eq.\,\eqref{ham:kmh-kspace}), we call the new operators $u_{\bs{k}\sigma}$ and $l_{\bs{k}\sigma}$ where $u$ and $l$ refer to upper and lower band.

\begin{figure}[t!]
\centering
\includegraphics[scale=0.55]{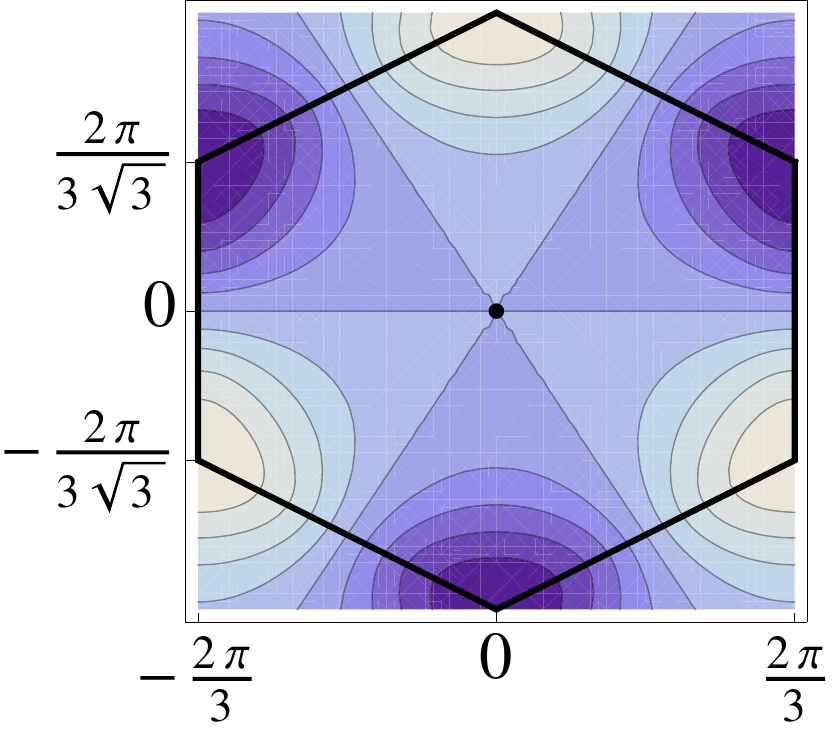}~~
\includegraphics[scale=0.46]{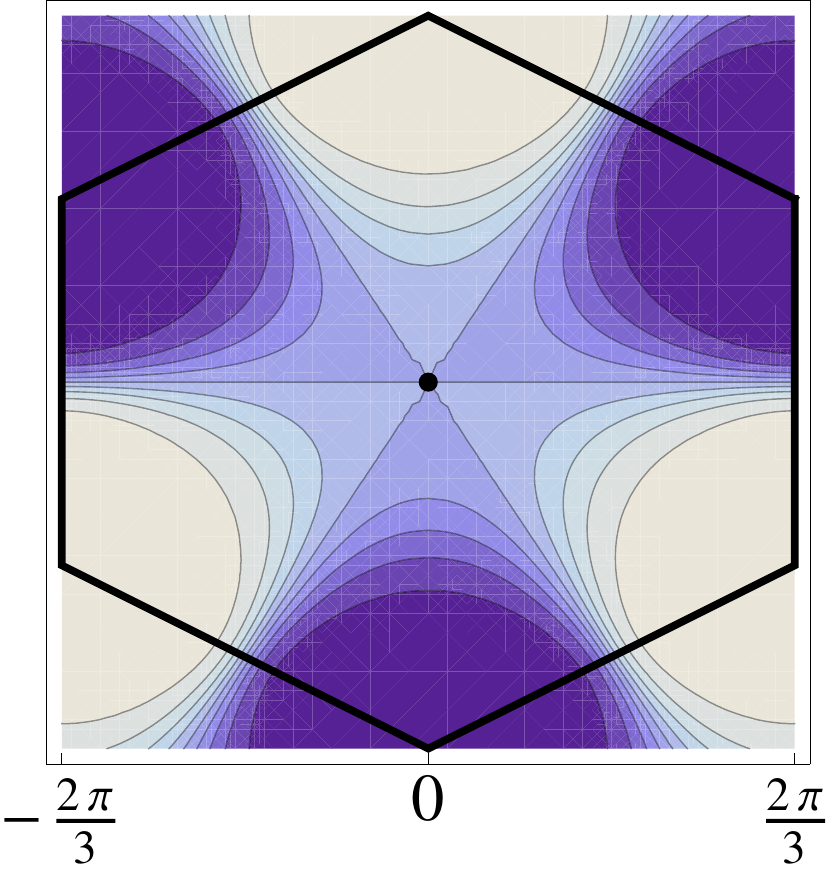}
\caption{(Color online) Exemplarily the function $|\alpha_-(\bs{k})|^2$ is shown in the BZ for $\lambda/t=0.2$ (left) and $\lambda/t=1.0$ (right).  The dark blue region corresponds to a value of zero, while the white region to a value of 1. The function $|\alpha_+(\bs{k})|^2$ is identical to $|\alpha_-(\bs{k})|^2$ by interchange of white and dark blue regions. The black lines mark the boundary of the BZ and the black dot in the center the $\Gamma$ point.}
\label{fig:A}
\end{figure}

The explicit transformation matrices between the sublattice basis $(a_{\bs{k}\sigma},b_{\bs{k}\sigma})$ and the band basis $(l_{\bs{k}\sigma},u_{\bs{k}\sigma})$ are given by
\begin{equation}
\left(\begin{array}{c}a_{\bs{k}\up}\\[0pt] b_{\bs{k}\up} \end{array}\right)=
\left( \begin{array}{cc} 
-\alpha_- & -\alpha_+ \\[0pt]
\beta_- & \beta_+ \end{array} \right)
\left(\begin{array}{c}l_{\bs{k}\up}\\[0pt] u_{\bs{k}\up} \end{array}\right)
\equiv T_\up \left(\begin{array}{c}l_{\bs{k}\up}\\[0pt] u_{\bs{k}\up} \end{array}\right)
\end{equation}
and
\begin{equation}
\left(\begin{array}{c}a_{\bs{k}\dw}\\[0pt] b_{\bs{k}\dw} \end{array}\right)=
\left( \begin{array}{cc} 
\alpha_+ & \alpha_- \\[0pt]
\beta_+ & \beta_-  \end{array} \right)
\left(\begin{array}{c}l_{\bs{k}\dw}\\[0pt] u_{\bs{k}\dw} \end{array}\right)
\equiv T_\dw \left(\begin{array}{c}l_{\bs{k}\dw}\\[0pt] u_{\bs{k}\dw} \end{array}\right) \ .
\end{equation}
We define the functions
\begin{eqnarray}\label{trafo:alpha}
\alpha_\pm &\equiv& \alpha_\pm(\bs{k})=\mathcal{N}_\pm d_\pm\ , \\[10pt]
\label{trafo:beta}
\beta_\pm &\equiv& \beta_\pm(\bs{k}) =\mathcal{N}_\pm \ ,
\end{eqnarray}
where 
\begin{equation}
d_\pm = \frac{g(\gamma \pm \eps)}{|g|^2}
\end{equation}
and
\begin{equation}
\mathcal{N}_\pm=
 \frac{|g|}{\sqrt{|g|^2+(\gamma\pm\varepsilon)^2}}\ .
\end{equation}
Note that $g$, $\gamma$, $\eps$ and then also $d_\pm$, $\alpha_\pm$, $\beta_\pm$, and $\mathcal{N}_\pm$ are $\bs{k}$-dependent. But for the sake of clarity we omit the 
$\bs{k}$-dependence. To give the reader a better idea about $\alpha_\pm$ and $\beta_\pm$, we have plotted $|\alpha_-(\bs{k})|^2$ for $\lambda/t=1.0$ and for $\lambda/t=0.2$ in Fig.\,\ref{fig:A}. 
Let us mention that $d_\pm$ is a complex function and hence $\alpha_\pm$, while $\beta_\pm$ is real. Nonetheless, to make the equations more symmetric, we will use the complex conjugate of $\beta_\pm$ as well.
Technical aspects and mathematical considerations
associated
with the change of basis are presented in Appendix\,\ref{sec:appA}.
There, we show useful formulas like $|\alpha_\pm|^2=|\beta_\mp|^2$ and obtain the important result:
\begin{equation}\label{sumalpha2}
\sum_{\bs{k}\in\,{\rm BZ}} |\alpha_\pm(\bs{k})|^2 =
\sum_{\bs{k}\in\,{\rm BZ}} |\beta_\pm(\bs{k})|^2 = \frac{N_\Lambda}{2}\ .
\end{equation}

Now we want to investigate the effect of the Hubbard term on the topological band insulator state more deeply. For that purpose, we transform the interaction term into the band basis.
Assuming $\lambda>0.2\,t$ we know from Eq.\,\eqref{kmh:gap-size} that the gap size $\Delta=2\,t$ is large and hence we can neglect all terms containing operators of the upper band (since it costs roughly the energy $\Delta$ to make any process between lower and upper band).
Then we decompose the remaining term in a standard way. Eventually the Hubbard term reduces to a chemical potential term, as we will see, where the chemical potential is given by $U/2$. The whole procedure does not use any additional assumptions and works for any interaction strength $U$.

\begin{equation}\begin{split}
&\quad\mathcal{H}_I' = U\sum_i n_{i\up}n_{i\dw} \\[5pt]
= & \frac{U}{N_\Lambda} \sum_{\bs{k}\bs{k}'\bs{q}}
a^\dagger_{\bs{k}+\bs{q}\up} a^\dagger_{\bs{k}'-\bs{q}\dw}
a^{\phantom{\dagger}}_{\bs{k}'\dw}a^{\phantom{\dagger}}_{\bs{k}\up}
+ b^\dagger_{\bs{k}+\bs{q}\up} b^\dagger_{\bs{k}'-\bs{q}\dw}
b^{\phantom{\dagger}}_{\bs{k}'\dw}b^{\phantom{\dagger}}_{\bs{k}\up} \\[5pt]
\approx&\frac{U}{N_\Lambda} \sum_{\bs{k}\bs{k}'\bs{q}}
\Big( \mathcal{A}(\bs{k},\bs{k}',\bs{q})\,+ \,\mathcal{B}(\bs{k},\bs{k}',\bs{q})\Big)
 l^\dagger_{\bs{k}+\bs{q}\up} l^\dagger_{\bs{k}'-\bs{q}\dw}
l^{\phantom{\dagger}}_{\bs{k}'\dw}l^{\phantom{\dagger}}_{\bs{k}\up} 
\end{split}
\end{equation}
where we have suppressed all terms containing operators $u_{\bs{k}\sigma}$ or $u_{\bs{k}\sigma}^\dagger$ in the last line.
The prefactors $\mathcal{A}$ and $\mathcal{B}$ are given by
\begin{equation}
\begin{split}
\mathcal{A}(\bs{k},\bs{k}',\bs{q})&=
\alpha^\star_-(\bs{k}+\bs{q}) \alpha^\star_+(\bs{k}'-\bs{q}) \alpha_+(\bs{k}') \alpha_-(\bs{k})\ , \\[10pt]
\mathcal{B}(\bs{k},\bs{k}',\bs{q})&=
\beta^\star_-(\bs{k}+\bs{q}) \beta^\star_+(\bs{k}'-\bs{q}) \beta_+(\bs{k}') \beta_-(\bs{k})\ .
\end{split}
\end{equation}
Now the Hubbard term will be decomposed as follows:
\begin{equation}
\begin{split}
& l^\dagger_{\bs{k}+\bs{q}\up} l^\dagger_{\bs{k}'-\bs{q}\dw}
l^{\phantom{\dagger}}_{\bs{k}'\dw}l^{\phantom{\dagger}}_{\bs{k}\up}
\approx
\delta_{\bs{q} 0}\langle l^\dagger_{\bs{k}\up} l^{\phantom{\dagger}}_{\bs{k}\up}\rangle
l^{{\dagger}}_{\bs{k}'\dw}l^{\phantom{\dagger}}_{\bs{k}'\dw}~+\\[5pt]
&\quad + \, l^\dagger_{\bs{k}\up} l^{\phantom{\dagger}}_{\bs{k}\up}
\delta_{\bs{q} 0} \langle l^{{\dagger}}_{\bs{k}'\dw}l^{\phantom{\dagger}}_{\bs{k}'\dw}\rangle
\,-\, \delta_{\bs{q} 0}\langle l^\dagger_{\bs{k}\up} l^{\phantom{\dagger}}_{\bs{k}\up}\rangle
\langle l^{{\dagger}}_{\bs{k}'\dw}l^{\phantom{\dagger}}_{\bs{k}'\dw}\rangle\ .
\end{split}
\end{equation}
Then the prefactors reduce to
\begin{equation}\begin{split}
\mathcal{A}(\bs{k},\bs{k}',\bs{q})\delta_{\bs{q} 0}=&|\alpha_-(\bs{k})|^2\,|\alpha_+(\bs{k}')|^2\ , \\[5pt]
\mathcal{B}(\bs{k},\bs{k}',\bs{q})\delta_{\bs{q} 0}=&|\beta_-(\bs{k})|^2\,|\beta_+(\bs{k}')|^2\ .
\end{split}
\end{equation}
We find the MF-decoupled interaction term in the new basis where we assume $\langle n_{\bs{k}\sigma}^l \rangle = \langle l_{\bs{k}\sigma}^\dagger l_{\bs{k}\sigma}^{\phantom{\dagger}}\rangle=1$  for $\sigma=\up, \dw$ since the lower band is completely filled.  Therefore, we get:
\begin{widetext}
\begin{equation}\label{eq-BHF}\begin{split}
\mathcal{H}_I' 
\approx&\sum_{\bs{k}\bs{k}'} \frac{U}{N_\Lambda}
\Bigg( \mathcal{A}(\bs{k},\bs{k}',0) n^l_{\bs{k}\dw}
+\mathcal{A}(\bs{k}',\bs{k},0) n^l_{\bs{k}\up} 
+\mathcal{B}(\bs{k},\bs{k}',0) n^l_{\bs{k}\dw}
+\mathcal{B}(\bs{k}',\bs{k},0) n^l_{\bs{k}\up} - 
 \mathcal{B}(\bs{k},\bs{k}',0)-\mathcal{A}(\bs{k},\bs{k}',0)  \Bigg)\\5pt]
=& U \sum_{\bs{k}} \Big(  |\alpha_-(\bs{k})|^2 \frac{1}{N_\Lambda}\sum_{\bs{k}'} 
|\alpha_+(\bs{k}')|^2 
n^l_{\bs{k}\up} + |\alpha_+(\bs{k})|^2  \frac{1}{N_\Lambda}\sum_{\bs{k}'} 
|\alpha_-(\bs{k}')|^2
n^l_{\bs{k}\dw} \Big. \\[3pt]
& \qquad\qquad\qquad\qquad\qquad \Big. + |\beta_-(\bs{k})|^2 \frac{1}{N_\Lambda}\sum_{\bs{k}'} 
|\beta_+(\bs{k}')|^2 n^l_{\bs{k}\up} + |\beta_+(\bs{k})|^2  \frac{1}{N_\Lambda}\sum_{\bs{k}'} |\beta_-(\bs{k}')|^2
n^l_{\bs{k}\dw}  \Big)~-~UN_\Lambda/2 \\[5pt]
=& \frac{U}{2} \sum_{\bs{k}}
\Big( \left(|\alpha_-(\bs{k})|^2 + |\beta_-(\bs{k})|^2\right) n^l_{\bs{k}\up} +
        \left(|\alpha_+(\bs{k})|^2 + |\beta_+(\bs{k})|^2 \right) n^l_{\bs{k}\dw} \Big) 
=~\sum_{\bs{k}}~\mu^{\rm eff} ~ n^l_{\bs{k}} \ ,
\end{split}
\end{equation}
\end{widetext}
with the effective chemical potential $\mu^{\rm eff}=U/2$ and $n^l_{\bs{k}}=n^l_{\bs{k}\up}
+ n^l_{\bs{k}\dw}$.
In the last line of Eq.\,\eqref{eq-BHF}, we have dropped the constant term.
For $\lambda>0.2t$, as the gap always spans from $-t$ to $t$, we conclude that as long as $U<2t$ the effective chemical potential lies in the gap, \ie the $U$ term does not affect the TBI gap. Thus we can assume that the physics in the KM model will be unchanged and consequently the edge modes persist (they will be described by the helical Luttinger liquid theory as a result of interactions \cite{wu-06prl106401}).

%

\subsection{General slave rotor arguments}
Now we will apply the slave rotor formalism which allows to address correlated electron systems at weak up to moderate interactions. The method has been introduced by Florens and Georges\cite{florens-02prb165111,florens-04prb035114}; we also refer the reader to the review by Zhao and Paramekanti\cite{zhao-07prb195101}.

Within this approach\cite{florens-02prb165111,florens-04prb035114,zhao-07prb195101}, the original fermion operators will be rewritten by a product of a fermionic operator $f_{i\sigma}$, the spinon or auxiliary fermion, and a phase factor $e^{i\theta_i}$, the rotor,
\begin{equation}
c_{i\sigma} = e^{i\theta_i}\,f_{i\sigma}\ .
\end{equation}
The idea is that the original fermions $c_{\sigma}$ are represented by a collective phase degree of freedom $\theta$ (conjugate to the total charge) and auxiliary fermions $f_{\sigma}$.
Introducing an additional variable, the angular momentum $L\propto i\pa_\theta$ associated with a quantum O(2) rotor $\theta$, simplifies then the original quartic interaction between the fermions as it is replaced by a simple kinetic term $\propto L^2$. State vectors in the new Hilbert space should have the form $\ket{\Psi}=\ket{\Psi_f}\ket{\Psi_\theta}$. The price we have to pay for the whole rewriting of the original problem is that the new Hilbert space is enlarged compared to the original one since unphysical states are present. To resolve this problem we have to impose a constraint,
\begin{equation}\label{global-constraint}
\sum_\sigma f_{i\sigma}^\dagger f_{i\sigma}^{\phantom{\dagger}} + L_i =1\ .
\end{equation}
Since the original fermion operators fulfill anticommutation relations also the spinon (or auxiliary fermion) operators do so. 
The reader may notice that in the rotor condensate phase the original electron and spinon operators are proportional, and one will find a situation where the spinon band structure describes physical electrons. In contrast, when the rotor is uncondensed, there is spin--charge separation and the spinons are emergent charge-neutral quasiparticles carrying spin only. The term ``spinon'' should not imply that the particles associated with the new $f$ operators obey fractional statistics in the spirit of the elementary excitation of the one-dimensional Heisenberg antiferromagnet. 

Rewriting the hopping term $\mathcal{H}_t$ yields
\begin{equation}\label{MF1}
\mathcal{H}_t = -t \sum_{\langle ij \rangle} \sum_\sigma 
\left( {f^a_{i\sigma}}^\dagger
{f^b_{j\sigma}}^{\phantom{\dagger}} e^{-i\theta_{ij}} + {\rm h.c.} \right)
\end{equation}
where $f^{a/b}$ refers to sublattice A/B and $\theta_{ij}\equiv \theta_i - \theta_j$. The spin orbit term $\mathcal{H}_{\rm SO}$  has a similar form,
\begin{equation}\label{MF2}
\mathcal{H}_{\rm SO} = i\lambda\sum_{\ll ij \gg}\sum_{\sigma\sigma'}
\nu_{ij} \sigma^z_{\sigma\sigma'}  f^{\dagger}_{i\sigma}f^{\phantom{\dagger}}_{j\sigma'} 
e^{-i\theta_{ij}}\ .
\end{equation}

Rewriting the Hubbard term \eqref{hubbard2} which is local yields
\begin{equation}\label{decomp-1}
\mathcal{H}_I = \frac{U}{2} \sum_i \left( \sum_\sigma n_{i\sigma}^f - 1 \right)^2 
= \frac{U}{2} \sum_i L_i^2\ ,
\end{equation}
where we have used the constraint and $n_{i\sigma}^f = f_{i\sigma}^\dagger f_{i\sigma}^{\phantom{\dagger}}$. 

Without proceeding further, the introduced formalism already allows to read off the following results. The Hubbard interaction $U$ affects the rotor sector only. As long as $U<t$ the rotors will condense favoring t 
 he uniform ansatz $\theta_{ij}=0$. It implies that the auxiliary fermions $f_{i\sigma}$ behave like the original electrons since $\exp{(\pm i\theta_{ij})}=1$ far away from the phase transition. Superfluid or Bose-condensed phases are known to be robust (roughly up to $U\sim t$). We can assume, hence, that the superfluid phase of the rotors also persists against moderate interactions before the phase transition in the insulating phase occurs. Both approaches presented in this section are beyond renormalization group (RG) arguments (the perturbative regime). Here, we have chosen to rewrite the Hubbard term in the rotor variables.

In the seminal paper by Kane and Mele\cite{kane-05prl226801} the stability of the TBI phase is briefly discussed. The derived RG equations indicate that additional Coulomb interactions increase the spin orbit gap size and does not destroy the TBI phase. The RG procedure, however, is only applicable in the perturbative regime where $t\gg U,\lambda$. In contrast, the arguments presented here provide evidence that for $\lambda > 0.2\,t$ the TBI phase is stable up to $U\sim t$ (against the Mott phase) reaching the strongly interacting regime. As a last point we shall mention that the shown stability of the TBI phase implies also the stability of the edge modes (which are described by the helical Luttinger liquid theory as a result of interactions \cite{wu-06prl106401}).

In Appendix\,\ref{sec:appB} we present the slave rotor approach using a simple approximation
to decompose Eq.\,\eqref{MF1}. We will restrict ourselves to the case $\lambda=0$.
When performing the mean field procedure it turns out that the used approximation leads to results which are not so reliable in finite dimensions. Therefore, in Sec.\,\ref{sec:SR}, we will apply the $\sigma$ model representation of the slave rotor theory to find the transition for finite spin orbit coupling where the rotors undergo a quantum phase transition from superfluid to Mott insulating phase. Before we address this issue, we briefly discuss the stability of the TBI phase in the context of $\mathbb{Z}_2$ invariants.

\subsection{$\mathbb{Z}_2$ invariants}

For the (non-interacting) KM model considered here, there exists in principle two non-trivial topological invariants, the spin-Chern number of Sheng \ea\cite{sheng-06prl036808} and the $\mathbb{Z}_2$ invariant proposed by Kane and Mele\cite{kane-05prl146802,kane-05prl226801}.

The spin-Chern number can be evaluated by integrating the Berry curvature of a fiber bundle obtained by imposing twisted boundary conditions\cite{sheng-06prl036808}. The procedure demonstrated in Ref.\,\onlinecite{sheng-06prl036808} implied that the spin Chern number is a robust topological invariant. Essentially the idea is that the system conserving $S^z$ decouples into two independent Hamiltonians for the up and down spins,
each Hamiltonian is characterized by a Chern integer. While the sum of the Chern integers is zero due to time reversal symmetry, its difference defines a quantized spin Hall conductivity\cite{sheng-06prl036808,hazan-10arXiv:1002.3895}. The $\mathbb{Z}_2$ invariant is then given by half the difference of the Chern integers (\ie the spin Chern number) modulo 2. Here we will choose another way and  calculate the $\mathbb{Z}_2$ invariant directly for the KM model. We follow Fu and Kane\cite{fu-07prb045302} and briefly adapt the calculation of the invariant for the sake of completeness.

\begin{figure}[t!]
\begin{center}
\includegraphics[scale=1.]{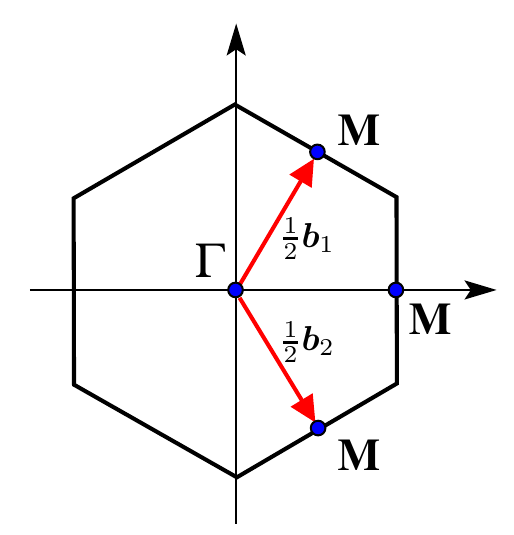}
\caption{(Color online) Brillouin zone with the four time-reversal invariant momenta $\Gamma$ and $M$. In addition, the reciprocal lattice vectors (with half of their length) are shown (red arrows).}
\label{fig:BZ-Z2}
\end{center}
\end{figure}
To compute the $\mathbb{Z}_2$ invariant it is important to keep the full tight binding model; in this section, we use the notations of Ref.\,\onlinecite{fu-07prb045302}. We express the matrix $\mathcal{H}_{\bs{k}}$ of Eq.\,\eqref{ham:kmh-kspace} in terms of $\Gamma$ matrices,
\begin{equation}
\mathcal{H}_{\bs{k}} = \sum_{a=1}^5 d_a(\bs{k})\Gamma^a
\end{equation}
where the five $\Gamma^a$ matrices are given by
$\Gamma^1=\tau^x\otimes I$, $\Gamma^2=\tau^y\otimes I$, $\Gamma^3=\tau^z\otimes\sigma^x$, $\Gamma^4=\tau^z\otimes\sigma^y$, and $\Gamma^5=\tau^z\otimes\sigma^z$. The coefficients $d_1$ and $d_2$ are essentially real and imaginary part of $g$ (more precise, real and imaginary part of $\exp{(-i\bs{k}\bs{\delta}_3)}\,g$), respectively, and $d_5=\gamma$. $d_3$ and $d_4$ are both zero. In addition to the five $\Gamma^a$ matrices, there are also their ten commutators $\Gamma^{ab}=[\Gamma^a,\Gamma^b]/(2i)$. Their anticommutators $\Gamma^a\Gamma^b+\Gamma^b\Gamma^a=2\delta_{ab}$ obey the Clifford algebra. The parity operator is defined as
\begin{equation}
\mathcal{P} = \tau^x \otimes I = \Gamma^1\ .
\end{equation}
Obviously, inversion $\mathcal{P}$ interchanges the sublattices ($\tau$) but not the spin ($\sigma$). The time-reversal operator $\mathcal{T}$ is defined by
\begin{equation}
\mathcal{T} = i ( I \otimes \sigma^y ) K\ ,
\end{equation}
where $K$ denotes complex conjugation. The Dirac matrices are chosen to be even under $\mathcal{P}\mathcal{T}$, $\mathcal{P}\mathcal{T}\,\Gamma^a(\mathcal{P}\mathcal{T})^{-1}=\Gamma^a$, while the commutators are odd under $\mathcal{P}\mathcal{T}$. Note that all $\Gamma^a$ are odd under $\mathcal{P}$ and $\mathcal{T}$ except $\Gamma^1$ which is even. Time-reversal and inversion symmetry imply that  the product $\mathcal{P}\mathcal{T}$ commutes with the Hamiltonian. The only time-reversal invariant points of the BZ which have to fulfill for a reciprocal lattice vector 
$-\bs{\Gamma}_i=\bs{\Gamma}_i+\bs{G}$ are given by
\begin{equation}
\bs{\Gamma}_i=\frac{1}{2}\left( n_1 \bs{b}_1 + n_2 \bs{b}_2 \right)
\end{equation}
with $n_i=0,1$.
We define $\bs{\Gamma}_1$ as the $\Gamma$ point of the BZ (\ie $\bs{k}=(0,0)$), $\bs{\Gamma}_2=\frac{1}{2}\bs{b}_1$, $\bs{\Gamma}_3=\frac{1}{2}\bs{b}_2$, and $\bs{\Gamma}_4=\frac{1}{2}(\bs{b}_1+\bs{b}_2)$ where the $\bs{b}_i$ are the reciprocal lattice vectors of the honeycomb lattice as shown in Fig.\,\ref{fig:BZ-Z2}. The latter three points are usually refered to as $M$ points. 
The $\mathbb{Z}_2$ invariants characterizing the occupied band are determined\cite{fu-07prb045302} by
\begin{equation}
\delta_i = -{\rm sign}\big( d_1(\bs{\Gamma}_i )\big)\ .
\end{equation}
The $\mathbb{Z}_2$ invariant $\nu=0,1$ which distinguishes a topological band insulator in two dimensions from a conventional band insulator is given by the product of all $\delta_i$,
\begin{equation}\label{Z2-inv}
(-1)^\nu = \prod_{i=1}^4 \delta_i\ .
\end{equation}
Since we can write $d_1(\bs{k})=t(1+\cos{(\bs{k}\bs{a}_1)} + \cos{(\bs{k}\bs{a}_2)})$ and use $\bs{a}_i\bs{b}_j =2\pi\delta_{ij}$, we find $d_1(\bs{\Gamma}_1)=3t$,  $d_1(\bs{\Gamma}_2)=d_1(\bs{\Gamma}_3)=t$, and $d_1(\bs{\Gamma}_4)=-t$. Thus $\delta_1=\delta_2=\delta_3=-1$ and $\delta_4=+1$ which implies by virtue of Eq.\,\eqref{Z2-inv}
\begin{equation}
\nu=1\ .
\end{equation}
In fact, a non-zero $\mathbb{Z}_2$ invariant implies a topological band insulator phase 
provided there is an energy gap throughout the BZ. In particular, the argument with invariants is a single-particle picture argument\cite{prodan09prb125327}. 

As the mean-field interacting Hamiltonian has been reduced to a single particle Hamiltonian (which leaves $d_1$ unchanged) the argument based on $\mathbb{Z}_2$ invariants is applicable as long as the spin orbit gap is present. The band Hartree Fock approach has precisely shown that the Hubbard term behaves as a chemical potential which lies between the bands as long as $|U|<2t$. Thus we have substantiated our earlier statement that the TBI phase and the presence of edge modes will be stable up to a region beyond the perturbative regime.

\section{Mott transition}\label{sec:SR}

Now we will use the slave rotor mean field
theory\cite{florens-02prb165111,florens-04prb035114} to find the transition where the charge degrees of freedom form a Mott insulating state (and not a band insulator). 

\subsection{Self-Consistency equations}
In contrast to the approximation in Appendix\,\ref{sec:appB} we will use a more sophisticated approach\,\cite{florens-02prb165111,florens-04prb035114,lee-05prl036403} in order to find the transition to the Mott phase. We replace the phase field representing the O(2) degree of freedom by a complex bosonic field $X=e^{i\theta}$ which is constrained by
\begin{equation}
|X(\tau)|^2=1\ .
\end{equation}
The associated Lagrange multiplier is called $\rho$. 
The derivation of the Lagrangian and the decomposition of hopping and spin orbit terms is shown in Appendix\,\ref{sec:appC}. The mean field parameters associated with the decomposition are given by
\begin{eqnarray}
\label{def:qx}
Q_X &=&\Big\langle\sum_\sigma {f_{i\sigma}^a}^\star {f_{j\sigma}^b}\Big\rangle\ , \\[5pt]
\label{def:qf}
Q_f &=& \Big\langle\exp{(-i\theta_{ij})}\Big\rangle\ ,
\end{eqnarray}
for the hopping term and
\begin{eqnarray}
\label{def:qxprime}
Q_X' &=& \Big\langle\sum_{\sigma\sigma'}  i \nu_{ij} \sigma_{\sigma\sigma'}^z f_{i\sigma}^\star f_{j\sigma'} \Big\rangle\ , \\[5pt]
\label{def:qfprime}
Q_f' &=& \Big\langle\exp{(-i\theta_{ij})} \Big\rangle\ ,
\end{eqnarray}
for the spin-orbit term. 
Finally we find the imaginary time Green's function for the 
 $f^l_\sigma$ fields, 
\begin{equation}\label{green:fl}
G_{fl} = \frac{1}{i\omega_n - \Sigma_{\bs{k}}}\ ,
\end{equation}
and for the $X$ fields,
\begin{equation}\label{green:X}
G_X =  \frac{1}{\nu_n^2/U + \rho + \xi_{\bs{k}}}\ .
\end{equation}
Here we introduced the renormalized KM spectrum for the spinon sector,
\begin{equation}\label{renorm-km-spec}
\Sigma_{\bs{k}} = \sqrt{\big( Q_f\,|g|\big)^2 + \big({Q_f'}\, \gamma\big)^2}\ ,
\end{equation}
and defined 
\begin{equation}\label{def:xi_k}
\xi_{\bs{k}} = -Q_X |g(\bs{k})| + Q_X'\lambda\,g_2(\bs{k})\ ,
\end{equation}
with $g_2$ from Eq.\,\eqref{g2}. 
Note that we consider only the half filled case here which allowed us to set $\mu=h=0$.
In the absence of spin orbit coupling, $\lambda\to 0$, we find $\Sigma_{\bs{k}} \to Q_f\,|g|$ and $\xi_{\bs{k}} \to -Q_X\,|g|$ and recover, hence, the Green's function from Florens and Georges\cite{florens-04prb035114} (when setting $\epsilon\equiv -|g|$). From there, we find directly the five self-consistency equations.
\begin{equation}\label{SCE1}
\begin{split}
1 &= \frac{1}{N_\Lambda}\sumk \frac{1}{\beta}\sum_n G_X(\bs{k},i\nu_n) \\[10pt]
&=\frac{U}{N_\Lambda}\sumk \frac{1}{\sqrt{ \Delta_g^2 +4U(\xi - \min{(\xi_{\bs{k}})})}}\ .
\end{split}
\end{equation}
In Eq.\,\eqref{SCE1} we performed the evaluation of the Matsubara sum at zero temperature (see Appendix\,\ref{sec:appD}) and we introduced the insulating gap
\begin{equation}
\Delta_g = 2\sqrt{U(\rho + \min{(\xi_{\bs{k}})})}\ .
\end{equation}
\begin{figure}[t!]
\centering
\includegraphics[scale=1.]{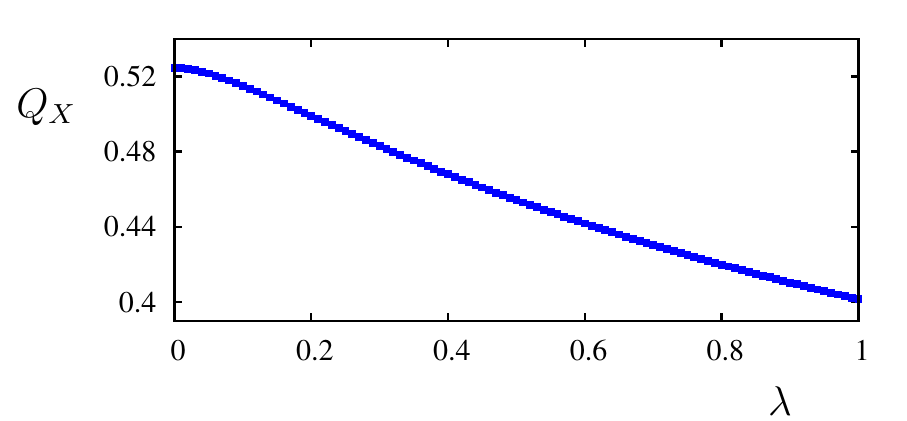}\\
\caption{(Color online) Numerical solution of the mean field equation \eqref{SCE2}. The behavior of $Q_X(\lambda)$ is shown ($t=1$).}
\label{fig:qx}
\end{figure}
While $\Delta_g$ is non-zero in the insulating phase, directly at the phase transition it will vanish since the rotors condense. Before we can use Eq.\,\eqref{SCE1} to find the transition line, we have to know the explicit form of $\xi_k$ and hence of $Q_X$ and $Q_X'$. The latter two mean field parameters are determined by use of the (second and third) self-consistency equations. We start with $Q_X$:
\begin{equation}\begin{split}
&t\sum_{j=1}^3 \sum_\sigma \langle {f_{i\sigma}^a}^\dagger {f_{j\sigma}^b} \rangle 
= \frac{1}{N_\Lambda} \sum_{\bs{k}\sigma} g(\bs{k}) \langle {f_{\bs{k}\sigma}^a}^\dagger {f_{\bs{k}\sigma}^b}\rangle \\[10pt]
&= \frac{1}{N_\Lambda}\sumk g \left( -\alpha_-^\star \beta_- + \alpha_+^\star \beta_+ \right)
\langle {f_{\bs{k}\sigma}^l}^\dagger {f_{\bs{k}\sigma}^l} \rangle 
\\[10pt]
&=\frac{1}{N_\Lambda}\sumk  \frac{|g|^2}{\sqrt{|g|^2 + \gamma^2}}\ .
\end{split}
\end{equation}
Due to the lattice symmetry, the sum over the three nearest neighbors, $\sum_{j=1}^3$, just 
appears as a factor 3 in the final expression. Thus we find the mean field parameter
\begin{equation}\label{SCE2}
Q_X = \frac{1}{3tN_\Lambda} \sumk  \frac{|g|^2}{\sqrt{|g|^2 + \gamma^2}}\ .
\end{equation}
We have plotted $Q_X$ as a function of $\lambda$ in Fig.\,\ref{fig:qx}.

\begin{figure}[t!]
\centering
\includegraphics[scale=0.95]{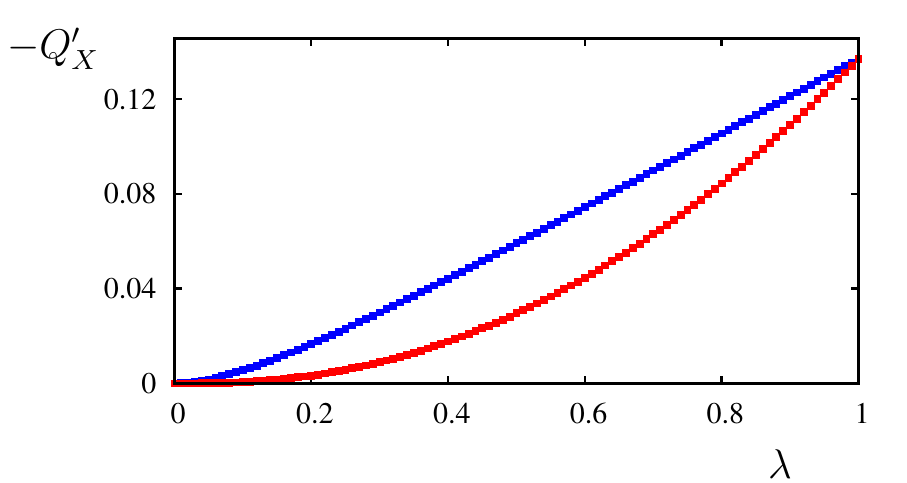}
\caption{(Color online) Numerical solution of $-\lambda Q_X'$ (red curve) and $-Q_X'$ (blue curve)  as a function of $\lambda$ ($t=1$) is shown.}
\label{fig:qxprime}
\end{figure}

In a similar way we proceed in order to find $Q_X'$:
\begin{eqnarray}
\nn\lambda\sum_{j=1}^6 Q_X' &=& \left\langle \lambda 
\sum_{j}\sum_{\sigma\sigma'} \,i\, \nu_{ij} \,\sigma_{\sigma\sigma'}^z 
f_{i\sigma}^\star f_{j\sigma} \right\rangle \\[10pt]
\nn&=&\frac{1}{N_\Lambda}\sumk \gamma \left\langle \Psi_{\bs{k}}^\dagger \sigma^z \tau^z \Psi_{\bs{k}}\right\rangle \\[10pt]
\nn&=&\frac{1}{N_\Lambda}\sumk \gamma \left( |\alpha_-|^2 - |\alpha_+|^2 - |\beta_-|^2 + |\beta_+|^2 \right)
\\[10pt]
&=& -\frac{1}{N_\Lambda}\sumk \frac{2\gamma^2}{\sqrt{|g|^2+\gamma^2}}\ .
\end{eqnarray}
Again the lattice symmetry is responsible for the fact that the sum over the next-nearest neighbors, 
$\sum_{j=1}^6$, can be replaced by a factor 6. Then the self-consistency equation reads
\begin{equation}\label{SCE2p}
\begin{split}
Q_X' =&  \left\langle \sum_{\sigma\sigma'} \,i\, \nu_{ij} \,\sigma_{\sigma\sigma'}^z 
f_{i\sigma}^\star f_{j\sigma} \right\rangle \\[10pt]
=&-\frac{1}{3\lambda N_\Lambda}
\sumk \frac{\gamma^2}{\sqrt{|g|^2+\gamma^2}} = -|Q_X'(\lambda)|\ .
\end{split}
\end{equation}
We have plotted $Q_X'$ as a function of $\lambda$ in Fig.\,\ref{fig:qxprime}.
With the knowledge of $Q_X$ and $Q_X'$ finally the rotor spectrum $\xi_{\bs{k}}$ of Eq.\,\eqref{def:xi_k} is well defined and we can proceed with Eq.\,\eqref{SCE1}. If one moves towards the transition from the Mott insulator to the superfluid phase of the rotors, the rotor gap $\Delta_g$ must close. It yields
\begin{equation}\label{Uc-SOI}
U_c(\lambda) = \left[ \frac{1}{2N_\Lambda}\sum_{\bs{k}'} \frac{1}{\sqrt{ \xi_{\bs{k}} - \min(\xi_{\bs{k}}) }} \right]^{-2}\ ,
\end{equation}
which defines the transition line between TBI and the EMI phase as shown in
the phase diagram Fig.\,\ref{fig:phasedia-sr}. The sum over $\bs{k}'$ means that formally the lowest bound corresponds to $\bs{k}\to\bs{k}_{\rm min}+\eta$, $\eta\ll 1$, and $\bs{k}_{\rm min}$ is associated with the minimum of $\xi_{\bs{k}}$. Hence, no divergence appears in the sum. A formal justification to cut the sum can be given by switching to ``energy space'' and considering the density of states. The same line of argument applies to Eqs.\,\eqref{qfc} and \eqref{qfcprime}.

\begin{figure}[t!]
\begin{center}
\includegraphics[scale=0.92]{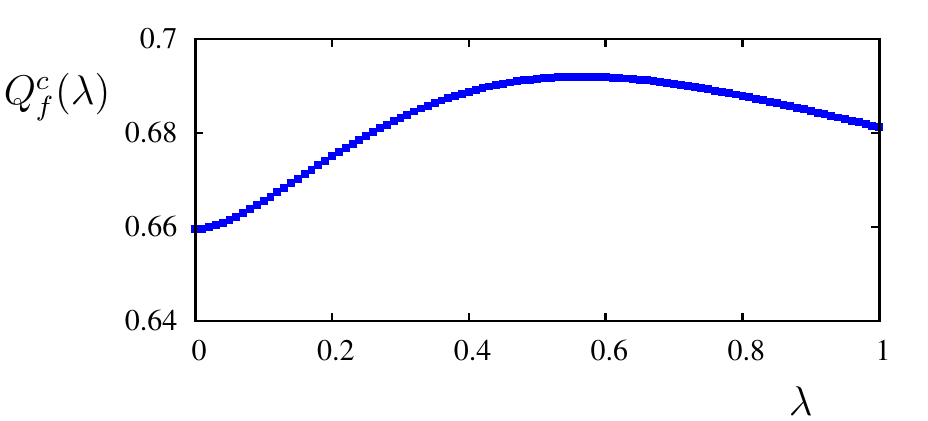}
\caption{(Color online) Numerical solution of $Q_f^c(\lambda)$, \ie $Q_f$ along the line $U_c(\lambda)$, as a function of $\lambda$ ($t=1$) is shown.}
\label{fig:Qfc}
\end{center}
\end{figure}

As a last point we have to consider $Q_f$ and $Q_f'$ and its behavior along the line 
$U_c(\lambda)$. Applying the same line of reasoning as for $Q_X$ we directly find
\begin{eqnarray}
\label{SCE3}
\nn Q_f &=& \langle X_i^\star X_j \rangle \Big|_{ij\,{\rm nn.}} = 
\frac{1}{N_\Lambda}\sumk e^{i\bs{k}\bs{\delta}_\mu} \langle 
{X_{\bs{k}}^a}^\star {X_{\bs{k}}^b} \rangle\\[10pt]
\nn &=& \frac{1}{N_\Lambda}\sumk \frac{|g|}{6t}\frac{1}{\beta}\sum_n G_X(\bs{k},i\nu_n)
\\[10pt]
&=&\frac{1}{N_\Lambda}\sumk \frac{|g|}{6t}\frac{U}{2\sqrt{U\left(\rho + \xi_{\bs{k}}\right)}}\ .
\end{eqnarray}
Here $\bs{\delta}_\mu$ denotes one of the three nearest-neighbor vectors. Along the transition line we have $\Delta_g=0$ and obtain
\begin{equation}\label{qfc}
Q_f^c(\lambda) = \frac{\sqrt{U_c(\lambda)}}{6tN_\Lambda}\sum_{\bs{k}'} 
\frac{|g|}{\sqrt{\xi_{\bs{k}}  
- \min{(\xi_{\bs{k}})}}} \ .
\end{equation}
It turns out that $Q_f^c$ is a slowly varying function of $\lambda$. We have plotted it in Fig.\,\eqref{fig:Qfc}.

\begin{figure}[t!]
\begin{center}
\includegraphics[scale=0.92]{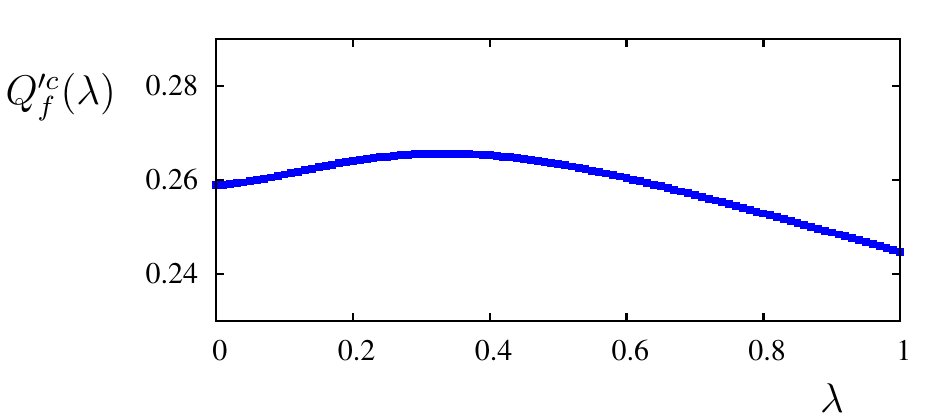}
\caption{(Color online) Numerical solution of $Q_f'^c(\lambda)$ as a function of $\lambda$ ($t=1$) is shown.}
\label{fig:qfcprime}
\end{center}
\end{figure}

The last self-consistency equation determines $Q_f'$.
\begin{eqnarray}
\label{SCE3p}
\nn Q_f' &=& \langle X_i^\star X_j \rangle \big|_{ij\,{\rm nnn.}}  \\[10pt]
\nn &=& \frac{1}{N_\Lambda}\sumk e^{i\bs{k}\bs{\delta}'_\mu} \langle {X_{\bs{k}}^{(a/b)}}^\star {X_{\bs{k}}^{(a/b)}} \rangle\\[10pt]
\nn &=&
\frac{1}{N_\Lambda}\sumk e^{i\bs{k}\bs{\delta}'_\mu} \frac{1}{2\beta}\sum_n
G_X(\bs{k},i\nu_n) \\[10pt]
&=& \frac{U}{2N_\Lambda}\sumk e^{i\bs{k}\bs{\delta}'_\mu}\frac{1}{2\sqrt{U(\rho+\xi_{\bs{k}})}}\ ,
\end{eqnarray}
with $\bs{\delta}'_\mu$ being one of the six next nearest neighbor vectors.
Thus we find $Q_f'$ along the Mott transition,
\begin{equation}\label{qfcprime}
Q_f'^c(\lambda) = \frac{\sqrt{U_c(\lambda)}}{2N_\Lambda}\sum_{\bs{k}'}\frac{e^{i\bs{k}\bs{\delta}'_\mu}}{\sqrt{\xi_{\bs{k}} 
 - \min{(\xi_{\bs{k}})}}}\ ,
\end{equation}
which we have plotted in Fig.\,\ref{fig:qfcprime}. 
From Figs.\,\ref{fig:Qfc} and \ref{fig:qfcprime} we see that  $Q_f^c$ and $Q_f'^c$ behave similarly but $Q_f^c$ is (roughly) three times larger than $Q_f'^c$. 

\subsection{Discussion}

\begin{figure}[t!]
\begin{center}
\includegraphics[scale=1.]{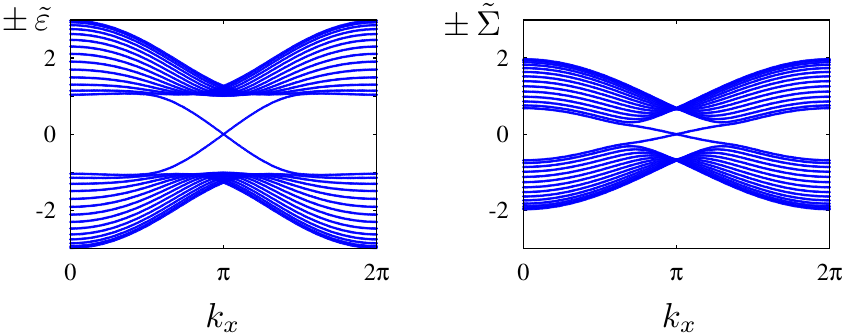}
\caption{(Color online) Left: Spectrum $\pm \, \tilde \eps(k_x)$ of the KM model on a stripe geometry as explained in the text for $\lambda=0.2$ and $t=1$. Right: Spectrum $\pm \, \tilde \Sigma(k_x)$ of the spinon sector for the same parameters at the critical line $U_c(\lambda=0.2)$. The hopping and spin orbit amplitudes are renormalized with $Q_f^c=0.68$ and $Q_f'^c=0.26$. The tildes refer to the finite stripe geometry in contrast to the spectra of the infinite system.}
\label{fig:spec-lso02}
\end{center}
\end{figure}

In Fig.\,\ref{fig:spec-lso02}, we show the spectrum of the non-interacting KM model for $\lambda=0.2$ as well as the renormalized spinon spectrum for $U_c(\lambda=0.2)$.
The geometry we considered is a stripe with 14 unit cells in $y$-direction while the stripe is infinitely long in $x$-direction.
Here we see that interactions contribute through $Q_f$ and $Q_f'$ such that the bulk spin gap is decreased compared to the TBI phase. 

At the {\it mean-field} level, spin-charge separation will occur and, while the charge is frozen in the Mott insulating state, the spin degrees of freedom will exhibit an Hamiltonian reminiscent of the KM model.
In particular, this implies the existence of gapless edge spinons.
In this sense this gives rise to the Fractionalized TI mentioned in the introduction\cite{young-08prb125316} or a ``topological Mott insulator''  as Pesin and Balents\cite{pesin-09np376} did.
Note that the ``topological Mott insulator'' phase has a different meaning than in the work of Raghu~\ea\cite{raghu-08prl156401} where the topological band insulator phase was caused by strong interactions. 
On the other hand, as already mentioned in the introduction, U(1) gauge fields associated with the lattice theory\cite{lee-05prl036403}, see also Appendix\,\ref{sec:appE},  cannot be ignored
in two dimensions. In particular, to stabilize the Fractionalized TI beyond the mean-field level, one
requires extra layers supporting gapless spinons allowing to screen the gauge field\cite{young-08prb125316} . (The stability of spinon excitations at the edges results from the fact that
single-particle tunneling is suppressed as a result of Mott physics \cite{lehur01prb165110}). This also implies that
in the context of an isolated (single) honeycomb layer the Fractionalized TI is unstable to instanton
proliferation and to easy plane Neel ordering\cite{ran-08arxiv:0806.2321,hermele-08prb224413}. 

Let us assume that conditions are realized such that the Fractionalized TI is stable against gauge fluctuations. Then, we can be even more precise when focusing on the spinon bulk sector. We can write for the corresponding ground state wave function,
\begin{equation}
\ket{\Psi_f} = \prod_{\bs{k}\sigma=\up,\dw} {f_{\bs{k}\sigma}^l}^\dagger \vac\ ,
\end{equation}
\ie the lower band $-\Sigma_{\bs{k}}$ (see Eq.\,\eqref{renorm-km-spec}) is completely filled. The explicit knowledge of $\ket{\Psi_f}$ allows us
to calculate the expectation value of the $z$-component of spin,
\begin{eqnarray}
\nn  && \bra{\Psi_f} \frac{1}{2}\left( n^f_{i\up} - 
n^f_{i\dw} \right) \ket{\Psi_f}\\[4pt]
\nn &=& \frac{1}{2 N_\Lambda} \sum_{\bs{k}} \bra{\Psi_f}
\left( |\alpha_-|^2 {f^{l}_{{\bs{k}}\up}}^\dagger {f^{l}_{{\bs{k}}\up}}^{\phantom{\dagger}} -
|\alpha_+|^2{f^{l}_{{\bs{k}}\dw}}^\dagger {f^{l}_{{\bs{k}}\dw}}^{\phantom{\dagger}} \!\!
\right) \ket{\Psi_f}\\[4pt]
&=& \frac{1}{2 N_\Lambda} \sum_{\bs{k}} 
\left( |\alpha_-|^2 - |\alpha_+|^2 \right) = 0\ .
\end{eqnarray}
In the last line we have used Eq.\,\eqref{sumalpha2}.
In the same way, we can easily check that $\langle S_i^x \rangle =0$.
At the mean-field level, we can also calculate the spin-spin correlation functions $\langle S_i^+ S_j^- \rangle$ for $i$ and $j$ on the same sublattice or on different sublattices. We show the former case explicitly:
\begin{eqnarray}
\langle S_i^+ S_j^- \rangle &=&
\bra{\Psi_f} {f_{i\up}^a}^\dagger {f_{i\dw}^a}^{\phantom{\dagger}}
{f_{j\dw}^a}^\dagger {f_{j\up}^a}^{\phantom{\dagger}} \!\!\! \ket{\Psi_f}\\[8pt]
\nn &=&\frac{1}{N_\Lambda^2}\sum_{\bs{k}_1,\bs{k}_2,\bs{k}_3,\bs{k}_4}
e^{-i\bs{k}_1\bs{R}_i+i\bs{k}_2\bs{R}_i-i\bs{k}_3\bs{R}_j + i\bs{k}_4\bs{R}_j}\\[3pt]
\nn &&\qquad \times
\big\langle {f_{\bs{k}_1\up}^a}^\dagger {f_{\bs{k}_2\dw}^a}^{\phantom{\dagger}}
{f_{\bs{k}_3\dw}^a}^\dagger {f_{\bs{k}_4\up}^a}^{\phantom{\dagger}} \!\! \big\rangle\\[8pt]
\nn &=&\left( \frac{1}{N_\Lambda}\sum_{\bs{k}_1}e^{-i\bs{k}_1(\bs{R}_i-\bs{R}_j)}
|\alpha_-(\bs{k}_1)|^2\right)\\[3pt]
\nn && \qquad \times \left( \frac{1}{N_\Lambda}\sum_{\bs{k}_2}e^{i\bs{k}_2(\bs{R}_i-\bs{R}_j)}
|\alpha_+(\bs{k}_2)|^2\right)\ ,
\end{eqnarray}
where we can evaluate the last line numerically and find that the correlations decay to zero on very short distances. In fact, when the distance $|\bs{R}_i - \bs{R}_j|$ reaches roughly four unit cells, the correlations are already smaller than $10^{-6}$. A similar calculation for both $i$ and $j$ on sublattice B as well as $i$ and $j$ on different sublattices reveals comparable results. We expect that the one-dimensional
character of spinon excitations at the edges gives rise to power-law spin correlation
functions. Other exotic spin liquid phases may be found in the vicinity of a Mott state\cite{lehur-09ap1452}.

The only ``hidden'' order which seems to be present is reflected in $\bra{\Psi_f}\sumk\gamma\, \Phi_{\bs{k}}^\dagger \sigma^z \tau^z \Phi_{\bs{k}} \ket{\Psi_f}\not= 0$ being a remnant of the TBI phase; here we have introduced the corresponding vector
$\Phi^\dagger_{\bs{k}}=\big( {f_{\bs{k}\up}^a}^\dagger, {f_{\bs{k}\up}^b}^\dagger,
{f_{\bs{k}\dw}^a}^\dagger, {f_{\bs{k}\dw}^b}^\dagger \big)$. 

In the phase diagram of Fig.\,\ref{fig:phasedia-sr} and \,\ref{fig:phasedia-sr1} the region above the TBI phase has to be handled with care for $0<\lambda<0.1\,t$ and is beyond the scope of this paper. 
In particular, in the absence of spin orbit coupling ($\lambda = 0$), mean-field slave rotor techniques predict a Mott insulator with gapless spin excitations\cite{lee-05prl036403}.
In the limit $\lambda\to 0$ we recover the earlier result of Lee and Lee\cite{lee-05prl036403}, \ie $U_c=1.68\,t$. 
On the other hand, in a recent QMC study it was shown\cite{meng-10n847} that the intermediate phase for $\lambda=0$ is an RVB spin liquid in contrast to Refs.\,\onlinecite{lee-05prl036403,hermele07prb035125}. 

While we have clarified the question what the effect of a Hubbard onsite interaction is one could also consider nearest and next nearest neighbor repulsion with amplitudes $V_1$ and $V_2$. Such a model in absence of spin orbit coupling was investigated by Raghu\,\ea\cite{raghu-08prl156401}. From the band Hartree Fock approach presented in Sec.\,\ref{sec:BHF} we see, however, that the effect for small $V_1$ and $V_2$  is negligible. This is in correspondence with Ref.\,\onlinecite{raghu-08prl156401} where strong nearest and next nearest neighbor interactions are required to reach QSH phases while weak interactions leave the semi metal unchanged.
As the intrinsic spin orbit interaction already opened a gap, $V_1$ and $V_2$ should be of the same order to affect the phase diagram. As large values of $V_1$ and $V_2$ are in the model under consideration somewhat unphysical, we can conclude that additional weak nearest and next nearest neighbor interactions are negligible for the KM model.

\section{Conclusion}\label{sec:conclusion}

We have investigated the Kane--Mele model in the presence of a Hubbard interaction as a paradigm for two--dimensional topological insulators with interactions. 

Using a mean-field procedure and arguments from the slave-rotor theory, we have shown that the TBI phase characterized by a $\mathbb{Z}_2$ topological invariant
is stable (against Mott phases) up to moderate interactions which are beyond the perturbative regime. 
The topological band insulator phase is separated from a Mott insulating region through $U_c(\lambda)$. 

At the mean-field level, charge constituents become frozen in the Mott state while the spin constituents form a quantum spin liquid with gapless edge spinons (preserving the time-reversal symmetry). Even though this Fractionalized TI phase is unstable against gauge fluctuations in the isolated honeycomb lattice
system\cite{ran-08arxiv:0806.2321,hermele-08prb224413}, the vicinity of other screening layers exhibiting gapless spinon excitations\cite{young-08prb125316} would allow to stabilize 
such a phase of matter in (quasi-) two dimensional systems. For very large onsite interactions, the Fractionalized TI phase inevitably turns into a SDW phase with XY ordering. 

For very weak spin orbit interactions, other insulating phases reminiscent of the ``gapless Mott insulator'' phase of Pesin and Balents\cite{pesin-09np376} might exist. 
It remains an open question if such a possible phase might be connected with the expected spin liquid phase for $\lambda\to 0$.

\begin{acknowledgements}
We acknowledge discussions with L. Balents, M. Hermele, Y.-B. Kim, J. Moore, G. Murthy, and R. Shankar. This work was supported by NSF Grant No.\ DMR-0803200 and by the Yale Center for
Quantum Information Physics (DMR-0653377). KLH also acknowledges support from CNRS at LPS Orsay, UMR 8502, and SR from the Deutsche Forschungsgemeinschaft (DFG) under Grant No.\ RA~1949/1-1.
\end{acknowledgements}

\appendix
\section{Band basis}\label{sec:appA}

In this Appendix, we present some supplementary material concerning the band basis which was introduced in Sec.\,\ref{sec:BHF}.
First, let us check, that $T_\up^\dagger T_\up=T_\up T_\up^\dagger = 1$ and $T_\dw^\dagger T_\dw=T_\dw T_\dw^\dagger = 1$.
We find the diagonal elements $|\alpha_\pm|^2+|\beta_\pm|^2 =1$ which can be easily checked. The off-diagonal elements are given by 
\begin{equation}\begin{split}
&\alpha_-^\star \alpha_+ + \beta_-^\star\beta_+ = \mathcal{N}_-\mathcal{N}_+
\big( d_-^\star d_+ + 1 \big) \\[5pt]
&\quad =  \mathcal{N}_-\mathcal{N}_+
\left( \frac{\gamma^2-\eps^2+|g|^2}{|g|^2} \right)=0\ .
\end{split}\end{equation}
We will further need the following expressions,
\begin{equation}
\mathcal{N}_+\mathcal{N}_- = \frac{|g|}{2\eps} \qquad\hbox{and}\qquad
\mathcal{N}_\pm^2 = \frac{|g|^2}{2(\eps^2 \pm \gamma\eps)}\ .
\end{equation}
In the above transformations the limit $g\to 0$ should be handled with care as the original eigenvectors diverge. This is a consequence of the fact that the matrix $\mathcal{H}_{\bs{k}}$ is already diagonal for $t=0$.
To consider the case with $\lambda=0$, we recover the transformation matrix \eqref{trafo:gamma=0} from Section\,\ref{sec:model},
\begin{equation}
\lim_{\gamma\to 0} T_\up = \lim_{\gamma\to 0} T_\dw = T_0\ .
\end{equation}
In order to prepare the following part of the Section, we have to show explicitly, that $\sumk |\alpha_\pm(\bs{k})|^2=\sumk |\beta_\pm(\bs{k})|^2=N_\Lambda/2$. 
One can show that $|\alpha_\pm|^2=|\beta_\mp|^2$ (which implies $|\alpha_+|^2+|\alpha_-|^2=1$) and it is sufficient to consider $|\alpha_\pm|^2$ in the following.
First we show that $|\alpha_\pm(\bs{k})|^2+|\alpha_\pm(-\bs{k})|^2=1$ and then we argue that the result follows directly. By using 
$|g(-\bs{k})|=|g(\bs{k})|$, $\eps(-\bs{k})=\eps(\bs{k})$, and $\gamma(-\bs{k})=-\gamma(\bs{k})$ we could calculate 
$|\alpha_\pm(\bs{k})|^2+|\alpha_\pm(-\bs{k})|^2=1$ explicitly. This is not necessary since the only thing we have to show is
\begin{equation}\label{alpha-relation}
|\alpha_-(\bs{k})|^2 = |\alpha_+(-\bs{k})|^2\ .
\end{equation}
By looking at the definition of $\alpha_\pm$  and using $\gamma(-\bs{k})=-\gamma(\bs{k})$, Eq.\,\eqref{alpha-relation} turns out to be correct.
Now we have to divide the BZ into two parts, \eg as follows: 
BZ=$\mathcal{K}_1 \cup \mathcal{K}_2$ with 
$\mathcal{K}_1= \{\bs{k}\in[-\frac{2\pi}{3},0]\times[-\frac{4\pi}{3\sqrt{3}},0]$ and 
$\bs{k}\in[0,\frac{2\pi}{3}]\times[-\frac{4\pi}{3\sqrt{3}},0[\,\}$ while $\mathcal{K}_2$ contains the remainder of the BZ such that $\mathcal{K}_1 \cap \mathcal{K}_2 =0$.
This ensures that $\bs{k}\in\mathcal{K}_1$ implies $-\bs{k}\in\mathcal{K}_2$ and vice versa.

Hence we can split the sum over the BZ as 
\begin{eqnarray}
\nn \sum_{\bs{k}\in {\rm BZ}}
 |\alpha_\pm(\bs{k})|^2 &=& \sum_{\bs{k}\in\mathcal{K}_1} |\alpha_\pm(\bs{k})|^2 + \sum_{\bs{k}\in\mathcal{K}_2} |\alpha_\pm(\bs{k})|^2\\
\nn  &=&\sum_{\bs{k}\in\mathcal{K}_1} \big( |\alpha_\pm(\bs{k})|^2 + |\alpha_\pm(-\bs{k})|^2 \big)\\
 &=& \sum_{\bs{k}\in\mathcal{K}_1} 1 = N_\Lambda/2\ .
 \end{eqnarray}

\section{Simple slave-rotor mean-field theory}\label{sec:appB}

In this Appendix, we perform a simple mean field decomposition for the KM model with Hubbard interactions which is rewritten in spinons and rotors. Starting from Eqs.\,\eqref{MF1} and \eqref{MF2}, adding the rewritten interaction term $H_I = U/2 \sum_i L_i^2$, we assume that state vectors in the Hilbert space should have the form $\ket{\Psi}=\ket{\Psi_f}\ket{\Psi_\theta}$. Decoupling the rotor and fermion variables and treating the constraint by introducing a Lagrange multiplier $h$, 
we obtain effective Hamiltonians for spinon and rotor sector:
\begin{eqnarray}
\nn \mathcal{H}^f &=& - \sum_{\langle ij \rangle\sigma} t_{ij}^{\rm eff} {f^a_{i\sigma}}^\dagger
{f^b_{j\sigma}}^{\phantom{\dagger}} + i
\sum_{\ll ij \gg}\lambda_{ij}^{\rm eff}\sum_{\sigma\sigma'}
\nu_{ij} \sigma^z_{\sigma\sigma'}  f^{\dagger}_{i\sigma}f^{\phantom{\dagger}}_{j\sigma'} \\
&&-(\mu + h)\sum_{i\sigma}f_{i\sigma}^\dagger f_{i\sigma}^{\phantom{\dagger}}\ , \\[10pt]
\nn \mathcal{H}^\theta &=& -\sum_{\langle ij \rangle} \mathcal{J}_{ij}^{\rm eff}\cos{(\theta_{ij})} +
\sum_{\ll ij \gg} \mathcal{G}_{ij}^{\rm eff} \cos{(\theta_{ij})} \\
&&+ \sum_i \left( \frac{U}{2}L_i^2 +hL_i\right)\ .
\end{eqnarray}
In the spinon sector we recover the original KM model expressed in $f_{\sigma}$ operators rather than $c_{\sigma}$ operators while the rotor sector corresponds to a quantum-XY like model for the phase variables. The effective amplitudes are determined by the following self-consistent equations,
\begin{eqnarray}
t_{ij}^{\rm eff} &=& t \left\langle \cos{(\theta_{ij})} \right\rangle_\theta\ , \\[10pt]
\lambda_{ij}^{\rm eff} &=& \lambda \left\langle \cos{(\theta_{ij})} \right\rangle_\theta\ , \\[10pt]
\mathcal{J}_{ij}^{\rm eff} &=&  t \sum_\sigma \left\langle {f_{i\sigma}^a}^\dagger
{f_{j\sigma}^b}^{\phantom{\dagger}} \right\rangle_f\ , \\[5pt]
\mathcal{G}_{ij}^{\rm eff} &=& \lambda \sum_{\sigma\sigma'} \left\langle i\nu_{ij} 
\sigma^z_{\sigma\sigma'}  f^{\dagger}_{i\sigma}f^{\phantom{\dagger}}_{j\sigma'} 
\right\rangle_f\ .
\end{eqnarray}
The expectation values are taken with respect to $\ket{\Psi_f}$ or $\ket{\Psi_\theta}$, respectively.
In the following, we show the main steps in finding the self consistency equations and the phase transition within this simple  approach. In fact, it is the analogous calculation for the honeycomb lattice to Sec.\ III.\,B.\,2.\ and 3.\ of Ref.\,\onlinecite{florens-04prb035114} 
for cubic lattices. We restrict ourselves to the case $\lambda=0$; the Hamiltonian $\mathcal{H}^f$ reads in momentum space 
\begin{equation}\label{Eq.B7}
 \mathcal{H}^f 
  = -\sum_{\bs{k}\sigma} \left( {f_{\bs{k}\sigma}^a}^\dagger,{f_{\bs{k}\sigma}^b}^\dagger \right)
\!\!  \left(\begin{array}{cc} \mu+h &  Z g \\[10pt]
   Z g^\star &  \mu+h \end{array}\right) \!\!
\left(\begin{array}{c} f_{\bs{k}\sigma}^a \\[10pt] f_{\bs{k}\sigma}^b  \end{array}\right)\ .
\end{equation}
Here we decomposed the expectation value simply as $\langle \cos{(\theta_i - \theta_j})\rangle \approx \langle \cos{\theta}\rangle^2=Z$.
By diagonalization of the matrix in Eq.\,\eqref{Eq.B7} we obtain the upper and lower band,
\begin{equation}
\epsilon_\pm = \pm Z |g| - (\mu+h)\ .
\end{equation}
We stress again that our constraint is differently chosen compared to Ref.\,\onlinecite{florens-04prb035114} where it is given by 
$\sum_\sigma f_{i\sigma}^\dagger f_{i\sigma}^{\phantom{\dagger}}  - L =1$. In 
the constraint used in this paper the $L$ term appears with a positive sign. The difference is due to the different definition of the fermions in term of spinons and rotors, see Eq.\,(6) of Ref.\,\onlinecite{florens-04prb035114}. Now, the Lagrange multiplier $h$ is defined from the following constraint equation:
\begin{equation}\label{constraint-average}
\langle L \rangle_\theta =  -\sum_\sigma \left( \langle f_{i\sigma}^\dagger f_{i\sigma}^{\phantom{\dagger}} \rangle_f - \frac{1}{2}\right)\ .
\end{equation}
To proceed further, we treat $\mathcal{H}^\theta$, which corresponds to a quantum XY model, at the mean-field level. The applied approximation, 
$\cos{(\theta_{ij})}\approx 2 \cos{(\theta_i)}\langle\cos{(\theta_j)}\rangle - {\rm const.}$, reduces the rotor Hamiltonian to a mean-field Hamiltonian of independent sites:
\begin{equation}
H^\theta_{\rm MF} = H_0 + H_I = \sum_i \left( \frac{U}{2} L_i^2 - h L_i \right) + \sum_i K \cos{\theta_i} \ .
\end{equation}
Here the coupling constant $K$ is given by
\begin{equation}
K=-2t\sum_\sigma\sum_j \langle {f_{i\sigma}^a}^\dagger
{f_{j\sigma}^b}^{\phantom{\dagger}} \rangle_f\langle \cos{\theta_j}\rangle_\theta \ .
\end{equation}
As long as we are in the rotor-condensed phase, we can assume $\langle \cos{\theta_j}\rangle \equiv \langle \cos{\theta} \rangle$ which allows us to evaluate $K$,
\begin{eqnarray}
\nn K&=& 2\langle \cos{\theta}\rangle \sum_\sigma \frac{1}{N_\Lambda}(-t)\sumk (g/t) \langle {f_{\bs{k}\sigma}^a}^\dagger {f_{\bs{k}\sigma}^b}^{\phantom{\dagger}} \rangle \\[10pt]
\label{B12} &=&4\langle \cos{\theta}\rangle \frac{1}{N_\Lambda}\sumk -g \frac{g^\star}{2|g|} \langle {f_{\bs{k}\sigma}^l}^\dagger {f_{\bs{k}\sigma}^l}^{\phantom{\dagger}} \rangle\\[10pt]
\nn &=& 2 \frac{1}{N_\Lambda} \sumk -|g|   \langle\cos{\theta}\rangle 
\equiv 2 \, \bar{\epsilon}(1/2) \,\langle\cos{\theta}\rangle\ .
\label{eps-bar}
\end{eqnarray}
The matrix elements in the previous equation are easily calculated using the definitions of Sec.\,\ref{sec:BHF},
\begin{equation}
\langle {f_{\bs{k}\sigma}^a}^\dagger {f_{\bs{k}\sigma}^b}^{\phantom{\dagger}} \rangle
= \left\{ \begin{array}{cccl}&-\alpha_-^\star \beta_- \langle  l_{\bs{k}\up}^\dagger l_{\bs{k}\up}^{\phantom{\dagger}} \rangle &\overset{\gamma\to 0}{=}&\frac{1}{2}\frac{g^\star}{|g|}\,\ ,
\\[10pt]
&\alpha_+^\star \beta_+ \langle  l_{\bs{k}\dw}^\dagger l_{\bs{k}\dw}^{\phantom{\dagger}}
\rangle
&\overset{\gamma\to 0}{=}&\frac{1}{2}\frac{g^\star}{|g|}\,\ .
\end{array}\right.
\end{equation}
We define the half-bandwith $D=3t$ for the honeycomb lattice, and find numerically the result 
\begin{equation}\label{1.57}
D\sum_\sigma \langle {f_{i\sigma}^a}^\dagger {f_{j\sigma}^b}^{\phantom{\dagger}} \rangle=|\bar\eps(1/2)| \simeq 1.57\ .
\end{equation}
Now let us consider Eq.\,\eqref{constraint-average},
\begin{equation}\begin{split}
\langle L \rangle =& -\sum_\sigma \left( \langle f_{i\sigma}^\dagger f_{i\sigma}^{\phantom{\dagger}} \rangle -\frac{1}{2} \right)  \\[10pt]
=&-\sum_\sigma \left( \! \frac{1}{N_\Lambda} \!\sumk \! \left( \frac{1}{2} 
\langle {f_{\bs{k}\sigma}^l}^\dagger
{f_{\bs{k}\sigma}^l}^{\phantom{\dagger}}\rangle +\frac{1}{2} \langle{f_{\bs{k}\sigma}^u}^\dagger {f_{\bs{k}\sigma}^u}^{\phantom{\dagger}}\rangle \right) \! -\frac{1}{2}\right)\\[10pt]
=&-2\left( \frac{1}{2} - \frac{1}{2} \right) =0\ ,
\end{split}
\end{equation}
where we assumed that the lower band is completely filled while the upper band is empty.
Also, since we still assume the half filled case, Eq.\,(40) of 
Ref.\,\onlinecite{florens-04prb035114} is easily
\begin{equation}
n=\frac{1}{2}\sum_\sigma \langle f_{i\sigma}^\dagger f_{i\sigma}^{\phantom{\dagger}}\rangle
=\frac{1}{2}\sum_\sigma \frac{1}{2} = \frac{1}{2}\ .
\end{equation}
This is equivalent to set $\mu+h=0$. Thus, we can also introduce $\mu_0(n)$
which is defined as:
\begin{equation}
\mu_0(n) = \frac{h+\mu}{Z} ~~\overset{{\rm half\,filling}}{\longrightarrow} ~~0\ .
\end{equation}
Similar to Ref.\,\onlinecite{florens-04prb035114} we obtain the Green's function
\begin{equation}
G_{fl}(\bs{k},i\omega_n)^{-1} = i\omega_n -Z\eps_{\bs{k}}\ .
\end{equation}
As a last step $\langle \cos{\theta}\rangle$ has to be calculated. Following Ref.\,\onlinecite{florens-04prb035114} we calculate it in first order perturbation theory in $K$. We start with
$\langle \cos{\theta} \rangle = \big\langle \psi_{l_n}^{(1)} \big| \cos{\theta} \big| \psi_{l_n}^{(1)} \big\rangle$ where
\begin{equation}
\big| \psi_{l_n}^{(1)} \big\rangle = \ket{ l_n}+\sum_{l\not= l_n} \frac{\bra{l}\cos{\theta}\ket{l_n}}{E_{l_n}-E_l}\ket{l}\ .
\end{equation}
To first order in $K$ only the ``mixed'' element contributes:
\begin{eqnarray}
\langle \cos{\theta} \rangle &=& 2K \sum_{l\not= l_n} \frac{| \bra{ l } \cos{\theta} 
\ket{l_n}|^2}{E_{l_n} - E_l }\ .
\end{eqnarray}
The energies are given by $E_l = 1/(2U) \big( UL+h\big)^2 + {\rm const.}$ and the matrix elements by $\frac{1}{4}|\bra{l}e^{i\theta} + e^{-i\theta}\ket{l_n}|^2=\frac{1}{4}(\delta_{l,l_{n-1}} +\delta_{l,l_{n+1}})$. Altogether we find the result
\begin{equation} 
\langle \cos{\theta}\rangle = -\frac{2K}{U}\ ,
\end{equation}
which is in agreement with Ref.\,\onlinecite{florens-04prb035114} at half filling.
Then we substitute $\langle \cos{\theta}\rangle$ in Eq.\,\eqref{B12} and
obtain finally
\begin{equation}
U_c = -4 \bar{\eps}(1/2) = 4 |\bar{\eps}(1/2)|\ .
\end{equation}
By means of Eq.\,\eqref{1.57} we find the phase transition at
\begin{equation}
U_c^\infty \simeq 6.30\,t\ , \label{Uc-infty}
\end{equation}
which should be considered as the correct result in $d=\infty$ dimensions. In this Appendix, we used a simple mean-field approximation with the severe restriction $Z=\langle\cos{(\theta_{ij})}\rangle$. This approximation might be justified in large
dimensions. Therefore, we can assume that the result $U_c^\infty$ is exact in $d=\infty$. 

\vspace{20pt}

\section{Derivation of Green's functions}\label{sec:appC}

In this Appendix, we pedagogically show all the relevant steps starting from the
slave rotor Hamiltonian to the Green's functions.
We have omitted this part in Sec.\,\ref{sec:SR} for the sake of clarity.
The slave rotor Hamiltonian reads
\begin{equation}\begin{split}
\mathcal{H}=&-t \sum_{\langle ij \rangle} \sum_\sigma 
\left( {f^a_{i\sigma}}^\dagger
{f^b_{j\sigma}}^{\phantom{\dagger}} e^{-i\theta_{ij}} + {\rm h.c.} \right) \\[7pt]
&+i\lambda\sum_{\ll ij \gg}\sum_{\sigma\sigma'}
\nu_{ij} \sigma^z_{\sigma\sigma'}  f^{\dagger}_{i\sigma}f^{\phantom{\dagger}}_{j\sigma'} 
e^{-i\theta_{ij}} \\[7pt]
&- \mu\sum_{i,\sigma} f_{i\sigma}^\dagger f_{i\sigma}^{\phantom{\dagger}} + \frac{U}{2}\sum_i L_i^2\ ,
\end{split}\end{equation}

where we still have to fulfill the constraint Eq.\,\eqref{global-constraint} with the Lagrange multiplier $h$. 
Then, the action is built from
\begin{equation}
S_0 \equiv \int_0^\beta d\tau \left[ -i L \pa_\tau\theta + \mathcal{H} + f^\dagger\pa_\tau f \right]\ ,
\end{equation}
where the first two terms correspond to the Legendre transform between $\mathcal{H}$
and $\mathcal{L}$ and we are switching from phase and angular momentum operator $(\theta,L)$ to fields $(\theta,\pa_\tau\theta)$. Here $L$ and $\pa_\tau\theta$ are related as follows,
\begin{equation}
i\pa_\tau \theta = \frac{\pa \mathcal{H}}{\pa L}\ ,
\end{equation}
which yields $L = (i/U)\,\pa_\tau\theta$.
We obtain the action
\begin{widetext}
\begin{equation}\label{fullaction}
\begin{split}
S_0=&\int_0^\beta d\tau \Bigg[ \sum_{i\sigma} f_{i\sigma}^\star \left( \pa_\tau - \mu + h_i \right)f_{i\sigma} + \frac{1}{2U}\sum_i\big( \pa_\tau\theta_i + ih_i \big)^2 \Bigg. \\[0pt]
\Bigg. &\qquad\qquad\qquad\qquad +\sum_i \left( -h_i + \frac{h_i^2}{2U}\right) -t\sum_{\langle ij \rangle}\left( \sum_\sigma {f_{i\sigma}^a}^\star {f_{j\sigma}^b} e^{-i\theta_{ij}} + {\rm c.c.}\right) + \lambda 
\sum_{\ll ij \gg}\left( \sum_{\sigma\sigma'} i\nu_{ij} \sigma^z_{\sigma\sigma'} 
f_{i\sigma}^\star f_{j\sigma} e^{-i\theta_{ij}}\right)
\Bigg]\ .
\end{split}
\end{equation}
\end{widetext}
Now we have the choice to decompose the hopping and spin orbit terms either in a standard way (as Florens and Georges did\cite{florens-04prb035114}) or in a more elaborate way (as Lee and Lee did\cite{lee-05prl036403}) to obtain an effective theory. Since we are mainly interested in the transition line to the Mott phase we restrict ourselves to the first way for the moment. Thus we will use again the decomposition $\alpha\beta\approx\langle\alpha\rangle\beta+\alpha\langle\beta\rangle-\langle\alpha\rangle\langle\beta\rangle$ 
with 
\begin{eqnarray}
\nn \alpha_{ij} =&\sum_\sigma {f_{i\sigma}^a}^\star {f_{j\sigma}^b},\quad\langle\alpha_{ij}\rangle &\equiv Q_X\ ,\\[5pt]
\nn \beta_{ij} =& \exp{(-i\theta_{ij})},\quad\quad \langle\beta_{ij}\rangle & \equiv Q_f\ ,
\end{eqnarray}
for the hopping term and
\begin{eqnarray}
\nn \alpha'_{ij} =& \sum_{\sigma\sigma'}  i \nu_{ij} \sigma_{\sigma\sigma'}^z f_{i\sigma}^\star f_{j\sigma'},~~ \langle\alpha'_{ij}\rangle & \equiv Q_X'\ , \\[5pt]
\nn \beta'_{ij} =& \exp{(-i\theta_{ij})},\qquad\qquad\quad \langle\beta'_{ij}\rangle & 
\equiv Q_f'\ ,
\end{eqnarray}
for the spin-orbit term. 
Then we replace the exponentials $\exp{(i\theta_i)}$ by complex bosonic fields $X(\tau)$
which are constraint via $|X_i|^2=1$. This constraint is imposed by a complex Lagrange multiplier $\rho_i$. Then the decomposed Lagrangian has the form 
\begin{equation}
S=S' + S'' + S'''\ ,
\end{equation}
where $S'$ contains the hopping term, $S''$ the spin orbit term, and $S'''$ the other terms. $S'$ is given by
\begin{equation}\begin{split}
S'\approx & \int_0^\beta d\tau \Big[ -|t| Q_X \sum_{\langle ij \rangle } X_i^\star X_j 
+ {\rm c.c.} \Big.\\[5pt]
\Big.&-|t| Q_f \sum_{\langle ij \rangle} \sum_\sigma f_{i\sigma}^\star f_{j\sigma} + {\rm c.c.} 
+ |t|\sum_{ij} Q_f Q_X \Big] \\[5pt]
=&\int_0^\beta d\tau \Big[ \mathcal{L}_X + \mathcal{L}_f + \ldots \Big]\ .
\end{split}\end{equation}
The second term $S''$ is given by
\begin{equation}\begin{split}
S''\approx & \int_0^\beta d\tau \Big[ \lambda Q_X'
\sum_{\langle\langle ij \rangle\rangle} X_i^\star X_j + {\rm c.c.} \Big. \\[5pt]
\Big.&+\lambda Q_f' \sum_{\langle\langle ij \rangle\rangle}
\sum_{\sigma\sigma'} \, i \nu_{ij} \sigma_{\sigma\sigma'}^z f_{i\sigma}^\star f_{j\sigma'}
+\lambda \sum_{\ll ij \gg} Q_f' Q_X' \Big] \\[5pt]
=&\int_0^\beta d\tau \left[ \mathcal{L}_X' + \mathcal{L}_f' +\ldots \right]\ .
\end{split}\end{equation}
Introducing the $X$-fields changes the Hubbard term such as 
\begin{equation}
\left( \pa_\tau\theta_i + h_i \right)^2
= \left[\left(i\pa_\tau + h_i\right)X_i^\star\right]\left[\left(-i\pa_\tau + h_i\right)X_i\right]\ ,
\end{equation}
and the term $S'''$ becomes the form
\begin{equation}\begin{split}
S''' =& \frac{1}{2U}\sum_i \left[\left(i\pa_\tau + h_i\right)X_i^\star\right]\left[\left(-i\pa_\tau + h_i\right)X_i\right]  \\[5pt]
& + \sum_i \rho_i |X_i|^2
+\sum_{i\sigma} f_{i\sigma}^\star \left( \pa_\tau - \mu + h_i \right)f_{i\sigma} +\ldots \\[5pt]
=& \int_0^\beta d\tau \left[ \mathcal{L}_X'' + \mathcal{L}_f'' + \ldots \right]\ .
\end{split}\end{equation}
Here and in the previous equations the ellipsis corresponds to the other terms which are independent of $f_\sigma$ and $X$.
The Fourier-transform of $\mathcal{L}_X$ and $\mathcal{L}_f$ leads in the $(u,l)$ basis to the bands obtained earlier:
\begin{eqnarray}
\nn \mathcal{L}_X &=& Q_X \sum_k (\,- g(\bs{k})\,){X_{\bs{k}}^a}^\star {X_{\bs{k}}^b} + {\rm c.c.} \\[5pt]
&=&Q_X\sum_k (\,-|g|\,){X_{\bs{k}}^l}^\star{X_{\bs{k}}^l} + |g|{X_{\bs{k}}^u}^\star{X_{\bs{k}}^u}\ ,\\[10pt]
\nn \mathcal{L}_f &=& Q_f \sum_{\bs{k}\sigma} (\,-g(\bs{k})\,) {f_{\bs{k}\sigma}^a}^\star
{f_{\bs{k}\sigma}^b} + {\rm c.c.} \\[5pt]
&=& Q_f \sum_{\bs{k}\sigma}  (\,-|g|\,) {f_{\bs{k}\sigma}^l}^\star
{f_{\bs{k}\sigma}^l} + |g|  {f_{\bs{k}\sigma}^u}^\star
{f_{\bs{k}\sigma}^u}\ .
\end{eqnarray}
For the $X$-part of the spin orbit term $\mathcal{L}_X'$ we find the following expression:
\begin{equation}
\begin{split}
\mathcal{L}_X' &= Q_X' \sum_{\bs{k}} \,\lambda \,g_2(\bs{k})\, \left( {X_{\bs{k}}^a}^\star {X_{\bs{k}}^a} +
{X_{\bs{k}}^b}^\star {X_{\bs{k}}^b} \right)\\[5pt]
&= Q_X' \sum_{\bs{k}} \,\lambda \,g_2(\bs{k})\, \left( {X_{\bs{k}}^l}^\star {X_{\bs{k}}^l} +
{X_{\bs{k}}^u}^\star {X_{\bs{k}}^u} \right)\ ,
\end{split}
\end{equation}
where $g_2(\bs{k})$ is the usual next-nearest neighbor hopping contribution as defined in Eq.\,\eqref{g2}.
The last term which must be transformed into momentum space is $\mathcal{L}_f'$
which clearly produces the $\gamma$-term. Therefore we will add $\mathcal{L}_f$ to $\mathcal{L}_f'$ in the $(f_\sigma^a,f_\sigma^b)$ basis and then transform both terms to the $(f_\sigma^l,f_\sigma^u)$ basis as we did with the original bands of the KM model:
\begin{equation}\begin{split}
\mathcal{L}_f+\mathcal{L}_f' =& \sum_{\bs{k}\sigma}
Q_f\left(-g(\bs{k}) {f_{\bs{k}\sigma}^a}^\star {f_{\bs{k}\sigma}^b} - g(\bs{k})^\star {f_{\bs{k}\sigma}^b}^\star {f_{\bs{k}\sigma}^a}\right) + \\[5pt]
&\sum_{\bs{k}\sigma\sigma'}\sigma^z_{\sigma\sigma'} Q_f'\,\gamma\, \left(
 {f_{\bs{k}\sigma}^a}^\star{f_{\bs{k}\sigma'}^a} -  {f_{\bs{k}\sigma}^b}^\star {f_{\bs{k}\sigma'}^b}\right)\\[10pt] 
=& \sum_{\bs{k}\sigma} -\Sigma_{\bs{k}} \,{f_{\bs{k}\sigma}^l}^\star {f_{\bs{k}\sigma}^l} 
 + \Sigma_{\bs{k}} \, {f_{\bs{k}\sigma}^u}^\star {f_{\bs{k}\sigma}^u} \ .
\end{split}\end{equation}
Here we have introduced the renormalized KM spectrum for the spinon sector,
\begin{equation} 
\Sigma_{\bs{k}} = \sqrt{\big( Q_f\,|g|\big)^2 + \big({Q_f'}\, \gamma\big)^2}\ .
\end{equation}
Finally we find the imaginary time Green's function for the 
 $f^l_\sigma$ fields, 
\begin{equation}
G_{fl} = \frac{1}{i\omega_n - \Sigma_{\bs{k}}}\ ,
\end{equation}
and for the $X$ fields,
\begin{equation}
G_X =  \frac{1}{\nu_n^2/U + \rho + \xi_{\bs{k}}}
\end{equation}
where we defined 
\begin{equation}
\xi_{\bs{k}} = -Q_X |g(\bs{k})| + Q_X'\lambda\,g_2(\bs{k})\ .
\end{equation}

\section{Matsubara Sum}\label{sec:appD}

In the self-consistency equations \eqref{SCE1}, \eqref{SCE3}, and \eqref{SCE3p} we had to evaluate the following Matsubara sum:
\begin{eqnarray}
\frac{1}{\beta}\sum_n G_X(\bs{k},i\nu_n)
&=& \frac{U}{\beta}\sum_n \frac{1}{\nu_n^2 + U(\rho+\xi_{\bs{k}})} \\[10pt]
\nn &=& \frac{U}{\beta}\sum_n \frac{1}{(i\nu_n +A)(-i\nu_n + A)}\ ,
\end{eqnarray}
where $A=\sqrt{U(\rho+\xi_{\bs{k}})}$. 
By taking the corresponding contour, we find
\begin{eqnarray}
\nn 0 &=& \oint_C \frac{n_B(z)}{(z+A)(z-A)} dz \\[10pt]
&=& 2\pi i \frac{1}{\beta} \sum_{n=0,\pm 1,\pm 2,\ldots} \frac{1}{(i\nu_n +A)(-i\nu_n + A)} \\[10pt]
\nn &&- 2 \pi i \left[ \frac{n_B(A)}{2A} - \frac{n_B(-A)}{2A}\right]\ ,
\end{eqnarray}
where $n_B(z)=( \exp{(\beta z)} -1 )^{-1}$ is the Bose function. The last equation then 
implies:
\begin{eqnarray}
\nn &&\frac{1}{\beta} \sum_{n\in\mathbb{Z}} \frac{1}{(i\nu_n +A)(-i\nu_n + A)}
=\frac{n_B(A) - n_B(-A)}{2A}\\[10pt]  
\nn &&= \frac{\coth{\left(\beta A/2\right)}}{2A} 
~~\overset{T\to 0}{\longrightarrow}~~ \frac{1}{2A}\ .
\end{eqnarray}
At zero temperature we find the result
\begin{equation}
\frac{1}{\beta}\sum_n G_X(\bs{k},i\nu_n) = \frac{U}{2\sqrt{U(\rho + \xi_{\bs{k}})}}\ .
\end{equation}

\vspace{20pt}
\section{Dynamical Gauge Field}\label{sec:appE}

In this Appendix, we decompose the action Eq.\,\eqref{fullaction} in a different way compared to Appendix\,\ref{sec:appC} and show, how the effective field theory and the gauge field $a_{ij}$ emerge.
Following Lee and Lee\,\cite{lee-05prl036403} we decompose the hopping term via Hubbard--Stratonovich decomposition, $\alpha_{ij}=\sum_\sigma {f_{i\sigma}^a}^\star {f_{j\sigma}^b}$ and $\beta_{ij}=e^{i\theta_{ij}}$,
\begin{eqnarray}
\nn&&\qquad\int d\eta_{ij} d\eta_{ij}^\star d\eta_{ji} d\eta_{ji}^\star \times \\[5pt]
\label{E1} &&\times e^{-\ell \left[ |\eta_{ij}|^2 + |\eta_{ji}|^2 - 
\eta_{ij}^\star \beta_{ij} - \eta_{ij}\alpha_{ij} - \eta_{ji}^\star\beta_{ji} - \eta_{ji}\alpha_{ji} 
\right]} \\[8pt]
\nn&&\quad = \frac{\pi^2}{\ell^2}\,e^{\ell\left[ \alpha_{ij}\beta_{ij}+\alpha_{ji}\beta_{ji}\right]}\ .
\end{eqnarray}
%
The same equation holds for $\alpha'$ and $\beta'$ in order to decouple the spin orbit term,
\begin{eqnarray}
\nn \alpha'_{ij} &=& \sum_{\sigma\sigma'}  i \nu_{ij} \sigma_{\sigma\sigma'}^z f_{i\sigma}^\star f_{j\sigma'}\\[5pt]
\nn \beta'_{ij} &=& \exp{(-i\theta_{ij})}.
\end{eqnarray}
In Eq.\,\eqref{E1}, $\ell$ is given by $\Delta \tau$ times the hopping or the spin orbit amplitude, respectively.
We further follow Lee and Lee and change the variables of integration by 
$\eta_{ij}=|\chi_{ij}|e^{-w_{ij} + i(a_{ij}^+ - a_{ij})}$ and $\eta_{ji}=|\chi_{ij}|e^{-w_{ij}+i(a_{ij}^+-a_{ij})}$. Note that $\eta_{ij}$ and $\eta_{ji}$ are independent complex variables, and hence $w_{ij}$ and $a_{ij}^+$ are independent and necessary.
At this point, we replace again $\exp{(i\theta_i)}$ by the bosonic $X_i$-field with the constraint $|X_i|^2=1$ imposed by the Lagrange multiplier $\rho_i$. 
Then we find the action which coincides with Eq.\,(4) of Ref.\,\onlinecite{lee-05prl036403} apart from two terms coming from the spin orbit interaction and the slightly different notation of the rotor variables. We replace the variables by their saddle point values plus fluctuations (see for details Ref.\,\onlinecite{lee-05prl036403}), neglect the massive modes and we can integrate out the $\rho_i$ field to restore the $\theta$ field.
Finally we obtain the effective Lagrangian (similar to Ref.\,\onlinecite{lee-05prl036403}): 
\begin{widetext}
\begin{equation}\label{eft}\begin{split}
L' =& \sum_{i\sigma} f_{i\sigma}^\star \left( \pa_\tau -i a_i^\tau + i\tilde h_i - \mu \right)
f_{i\sigma} + \frac{1}{2U} \sum_i \left( \pa_\tau \theta_i - a_i^\tau - \tilde h_i \right)^2
-\sum_{\langle ij \rangle, \sigma}|t| \tilde \chi_{ij}^X e^{i a_i^j} f_{j\sigma}^\star f_{i\sigma}
-\sum_{\langle ij \rangle} |t| \tilde \chi_{ij}^f e^{-(\theta_i - \theta_j - a_j^i)} \\[10pt]
&-\sum_{\ll ij \gg}\sum_{\sigma\sigma'} |\lambda | \tilde \chi_{ij}^{X'} e^{i a_i^j} i \nu_{ij} f_{j\sigma}^\star \sigma^z_{\sigma\sigma'} - \sum_{\ll ij \gg} |\lambda | \tilde \chi_{ij}^{f'} e^{-i(\theta_i - \theta_j - a_i^j)}\ .
\end{split}\end{equation}
\end{widetext}
Here $a_i^\tau$ and $a_i^j$ are the temporal and spatial gauge fields coming from the fluctuations from $h_i$ and $a_{ij}$, respectively. $h_i$ is the Lagrange multiplier associated with the global constraint introduced earlier. The quantities with tildes are the saddle point values and are identical to the mean-field parameters which we have evaluated in Sec.\,\ref{sec:SR}. We notice that both spinons and rotors couple to the U(1) gauge field.
Since we assume weak gauge fluctuations, we take the saddle point approximation, \ie $a_i^j=0$; we recover for the spinons the same terms which resulted in Sec.\,\ref{sec:SR} in the renormalized KM spectrum ($\tilde \chi_{ij}^X \to Q_X$ and $\tilde \chi_{ij}^{X'} \to Q_X'$). We should also mention that the spinons are still coupled to the gauge field through the first term in Eq.\,\eqref{eft} which contains $\sim f_{i\sigma}^\star a_i^\tau f_{i\sigma}$.
In principle, although we have set $a_i^j=0$, we could allow for small deviations and expand $\exp(i a_i^j)\approx 1 + i a_i^j$; thus the spinons couple to both temporal and spatial gauge fields.
On the other hand, we know that the rotors are gapped in the Mott phase and can be integrated out. This generates the Maxwellian term 
(see e.g. Refs.\,\onlinecite{ran-08arxiv:0806.2321}, \onlinecite{hermele-08prb224413}, \onlinecite{lee-05prl036403}).


\end{document}